\documentstyle[12pt,aaspp4]{article}

\begin{document}
\newbox\grsign \setbox\grsign=\hbox{$>$} \newdimen\grdimen \grdimen=\ht\grsign
\newbox\simlessbox \newbox\simgreatbox
\setbox\simgreatbox=\hbox{\raise.5ex\hbox{$>$}\llap
     {\lower.5ex\hbox{$\sim$}}}\ht1=\grdimen\dp1=0pt
\setbox\simlessbox=\hbox{\raise.5ex\hbox{$<$}\llap
     {\lower.5ex\hbox{$\sim$}}}\ht2=\grdimen\dp2=0pt
\def\gtorder{\mathrel{\copy\simgreatbox}}
\def\ltorder{\mathrel{\copy\simlessbox}}
\def\simgreat{\mathrel{\copy\simgreatbox}}
\def\simless{\mathrel{\copy\simlessbox}}

\def\degrees{^\circ }
\def\etal{{\it et al.} }
\def\cf{{\it cf.} }

\title{On the Distribution of Dust in the Large Magellanic Cloud$^1$}
\author{Jason Harris, Dennis Zaritsky,}
\affil{UCO/Lick Observatory and Department of Astronomy and
Astrophysics, Univ. of California, Santa Cruz, CA, 95064, E-Mail:
jharris@ucolick.org, dennis@ucolick.org}
\author{and}
\author{Ian Thompson}
\affil{Observatories of the Carnegie Institution of Washington,
813 Santa Barbara St., Pasadena, CA 91101, E-Mail: ian@ociw.edu}
\medskip
\vskip 3in
$^1$ Lick Bulletin No. 1365

\vfill
\eject

\abstract {We present a detailed map of the reddening in a 
$1.9\degrees \times 1.5\degrees$ section of the Large Magellanic Cloud
(LMC), constructed from $UBVI$ photometry of 2069 O and B main
sequence stars.  We use two reddening-free photometric parameters to
determine the line-of-sight reddening to these stars.  We find a mean
reddening, $\langle E(\bv) \rangle_{LMC} = 0.20$ mag, with a
non-Gaussian tail to high values.  When the reddening is corrected for
foreground Galactic extinction (\cite{ogs95}), we find $\langle E(\bv)
\rangle_{LMC} =0.13$ mag.   
The line-of-sight values are then interpolated onto a uniform 
grid with a local least-squares plane fitting routine to construct
a reddening map of the region.  We use the distribution of reddening
values to constrain the line-of-sight geometry of stars and dust in
the LMC, and to test and normalize a standard extinction correction
for galaxy photometry.  We attempt to distinguish between
line-of-sight depth effects and structure in the dust distribution as
possible causes for the observed differential reddening through this
region. 

We conclude: {\bf (1)} that our data are consistent with a vertical
exponential distribution of stars and dust in the LMC, for which the
dust scale height is twice that of the OB stars; {\bf (2)}
that the dust distribution must be non-uniform (clumpy) to account for 
the full distribution of measured reddening values ({\it i.e.},
line-of-sight effects alone are insufficient to explain the observed
structure); and {\bf (3)} that the $B$-band optical depth, $\tau_B$,
through the observed region of the LMC is $0.69 < \tau_B < 0.82$. }

\section{Introduction}\label{sec:intro}

Observations of galaxy properties, and therefore of the extragalactic
universe, are affected by internal dust extinction. Any correction for
this extinction is complicated by the unknown relative distributions
of dust and stars.  Empirical efforts to determine these distributions
have been based either on statistical analyses of multicolor images of
large samples of galaxies, or on detailed analysis of high spatial
resolution, multicolor images of a few, select nearby galaxies.
Theoretical efforts have depended upon the assumption of simple
geometries, even though the radiative transfer models are becoming 
more realistic with the inclusion of both scattering and absorption
(\cite{brz88}, \cite{wit92}).  Neither 
approach has succeeded in detail because of the complexity in
interpreting integrated galaxy properties.  The goal of this study is
to observationally constrain the star-dust geometry in one galaxy, the
Large Magellanic Cloud, for which we can observe lines-of-sight to
individual stars. 

Galaxy extinction corrections depend sensitively upon several factors,
including: {\bf (1)} the inclination of the galaxy (\cf \cite{val90},
\cite{bur91}, \cite{hva92}, \cite{gio94}); {\bf (2)} the stellar
population age and metallicity gradients (\cf \cite{dj96},
\cite{pel94}); and {\bf (3)} the details of the dust geometry 
({\it e.g.}, clumpiness; \cite{kch96}).  This sensitivity is reflected
by the wide range of results quoted in different studies (\cf
\cite{db95}). Disney (1989), Valentijn (1990) and Burstein \etal
(1991) have concluded that galactic disks are optically thick, even in
their outer regions.  Bosma (1992), Huizinga \& Van Albada (1992),
Byun (1993), Giovanelli \etal (1994) and others have found that the
optical depth is significant only near the center of disks. Further
study, particularly of a galaxy in which the stars can be individually 
resolved, is critical to determining the importance of clumpiness and
the relative distributions of the dust and different stellar
populations. 

The Magellanic Clouds are the most suitable galaxies for a
determination of the detailed internal distribution of dust and stars in
a galaxy and the subsequent effects on the observed properties.  
In this paper, we present a determination of the spatially-resolved
reddening in a section of the Large Magellanic Cloud.  Rather than
relying on integrated colors or model-dependent reddening
indicators, the line-of-sight reddening toward individual stars is
measured using four-filter photometry.  This allows us to avoid
complications related to scattering and stellar population gradients. 

This article is structured as follows.  In \S\ref{sec:data}, we
present the data and our reduction techniques.  In
\S\ref{sec:proc}, we discuss the construction of the reddening map,
including the determination of the line-of-sight reddening values, the
interpolation of these values to a uniform grid, and the results 
from several tests of the procedure.  In \S\ref{sec:discuss}, we discuss
the following: {\bf (1)} a comparison of our reddening values
to the results from previous studies of dust in the LMC, {\bf (2)} our 
constraint on the line-of-sight geometry of stars and dust in the
LMC, {\bf (3)} a derivation of the total $B$-band optical depth of the
LMC disk and a test of an extinction correction for the integrated
photometry of spiral galaxies, {\bf (4)} an investigation of whether 
coherent dust structure or incoherent effects, such as line-of-sight
depth differences, dominate the observed differential reddening across 
the LMC, and
{\bf (5)} a positionally-dependent reddening correction applied to the 
photometry of all stars in this section of the LMC.  We find that the
distribution of dust is non-uniform, and that it differentially affects
the photometry across the observed region.  This result appears
self-evident, but it is usually ignored in both empirical and
theoretical treatments of internal extinction.  Its inclusion
represents progress toward a greater understanding of the distribution
of stars and dust in the LMC. 

\section{Data}\label{sec:data}

The data, reduction, and photometric calibration are described in
detail elsewhere (\cite{zht97}), so we only briefly review their main 
characteristics here. The data come from the initial stages of a 
$UBVI$ photometric survey of the central $8\degrees \times 8\degrees$ 
of the LMC and $4\degrees \times 4\degrees$ of the SMC and were 
obtained using the Great Circle Drift Scanning Camera (\cite{zsb96})
at the Las Campanas 1m Swope telescope in November 1995.  The effective
exposure time is set by the sidereal drift rate of a star across the
field-of-view, which for this instrumental setup corresponds to about
4 minutes. The images cover a $1.9\degrees \times 1.5\degrees$ region
with 0.7 arcseconds per pixel resolution. The region is centered at
approximately $\alpha=5.2^h$ and $\delta=-67.4\degrees$. 

The data are reduced in a fairly standard manner using DAOPHOT II
(\cite{ste87}) and our own algorithms that automate the process as 
much as possible to produce a catalog of RA, Dec, and $U$, $B$,
$V$, and $I$ photometry for over one million stars. Four filter 
photometry is not available for every star because some stars are 
fainter than the respective magnitude limits in certain bands.  The
catalog becomes 50\% incomplete at $V\sim 21$ mag, well below the
apparent magnitude of OB stars ($V\simless 17$ mag).  We require that
a star is identified in both $B$ and $V$ to enter the
catalog. Photometric uncertainties are calculated by DAOPHOT and
convolved with uncertainties in the standard star calibration using
Landolt (1992) standard fields to produce our final photometric
uncertainties.  We find that this uncertainty at worst underestimates
the true uncertainty by a factor of two (\cf \cite{zht97}).

\section{Constructing the Reddening Map}\label{sec:proc}

The translation of stellar photometry into a reddening map involves
two steps.  First, we determine the line-of-sight reddening toward 
OB stars using two reddening-free combinations of the available 
photometric colors.  These data will be used to constrain
models of the dust geometry both along and across the line-of-sight.
Second, we interpolate the line-of-sight reddening values (LOSRV) onto a
rectilinear grid of points using a local least-squares plane fitting
routine to produce a two-dimensional map of the reddening.  We
describe the entire procedure below.

\subsection{Selection of Program Stars}\label{sec:select}

We use only OB main sequence stars to determine the LOSRV because 
{\bf (1)} they have small photometric errors due to their relative
brightness; {\bf (2)} their intrinsic colors have a negligible
dependence on metallicity (\cite{ogs95}), so we can use Galactic
stellar data to construct the relationship between reddening-free
parameters and intrinsic color, despite the mean metallicity
difference between the LMC and the Milky Way; and {\bf (3)} the
relationship between the intrinsic 
$\bv$ color, $(\bv)_0$, and the reddening-free parameter for these
stars is single-valued.  We select candidate OB stars using the
photometric criteria $V < 17 + 3.2 (\bv)$ and $\bv < 0.4$ (\cf Figure
\ref{fig:hess}).  The first criterion excludes main sequence stars
later than spectral type B5 (for $(m-M)_{LMC} = 18.47$; \cite{fw87}).
The color term in this expression anticipates the effects of
reddening by placing the boundary along a line of constant spectral
type, ensuring that faint stars with high reddening are not
under-represented, as discussed by Oestreicher and Schmidt-Kaler
(1996).  The second criterion excludes Galactic main sequence stars by
placing a cut blueward of the Galactic halo main sequence turn-off
(\cite{gil89}; \cf our Figure \ref{fig:hess}), but, unfortunately, may 
also exclude highly reddened ($E(\bv) > 0.7$ mag) LMC stars. We
discuss the number of highly obscured stars excluded by this 
criterion in \S\ref{sec:obsc}.  Finally, we exclude OB stars with
photometric uncertainties that lie beyond the half-maxima of the 
uncertainty distributions ($>$ 0.2, 0.1, 0.1, and 0.1 mag for $U$,
$B$, $V$, and $I$, respectively). 

As a final note about sample selection, we address the concern that
slightly evolved OB stars are photometrically degenerate with reddened
OB stars, and that these stars will cause us to overestimate the
reddening along their lines-of-sight.  Theoretical isochrones
(\cite{bbc94}) show that as an OB star evolves redward in $\bv$, it
becomes redder in $U-B$ in such a way that it remains
indistinguishable from an unreddened upper main sequence star. The
supergiant and main sequence loci diverge only redward of $\bv\sim 0$,
a situation that will require us to eliminate the more 
evolved supergiant stars from our sample (\cf \S\ref{sec:losrv}). 

\subsection{Determining the Line-of-Sight Reddening
Values}\label{sec:losrv}

We utilize two reddening-free photometric parameters to measure the
LOSRV,
$$Q_1=(U-B) - 0.76(\bv) - 0.05(\bv)^2$$
$$Q_2=(U-B) - 0.60(V-I) + X(V-I)^2.$$
These are equivalent to a rotation of the two-color diagram, so that
the reddening line is parallel to one axis.  $Q$ measures the
distance along the axis orthogonal to the reddening line; thus $Q$ is
reddening-free.  The quadratic terms correct for slight variations in
the slope of the reddening line for different spectral types.  $Q_1$
is a derivative of the familiar Johnson and Morgan (1953) parameter 
for the $UBV$ photometric system.  The coefficient of the linear 
$(\bv)$ term is the ratio of color excesses $E(U-B)/E(\bv)$ evaluated 
from the average LMC extinction curve outside 30 Dor (\cite{fit85}). 
The coefficient of the quadratic term is drawn directly from Galactic 
studies (\cite{hj56}). $Q_2$ is a hybrid of the Johnson and Morgan $Q$
and the $Q$ presented by Grieve and Madore (1986) for the $BVI$ system.  
The coefficient of the linear $(V-I)$ term is again the ratio of color
excesses $E(U-B)/E(V-I)$ evaluated from the average extinction curve
for the LMC outside 30 Dor.  To determine $X$ empirically, we
construct an artificial stellar sample, adopt a distribution of
reddening values consistent with our initial reddening estimates using
only $Q_1$ and the LMC extinction curve, and add random reddening
errors with $\sigma=0.04$ mag, consistent with the data.  We find $X$
to be $0.08$ by requiring $\langle Q_2 \rangle = \langle Q_1 \rangle$.
In practice, our final results are nearly independent of $X$ because a
star at the extreme red end of our color range has a quadratic  
term in $Q_1$ of 0.008, corresponding to a correction to $E(\bv)$
of only 0.002 magnitudes. A similar argument demonstrates that adopting
the Galactic value for the coefficient of the quadratic term in
$Q_1$ is acceptable. The quadratic terms only become important for
highly obscured stars, which we have excluded from the current sample.

The line-of-sight color excess, or reddening, is measured from a
comparison of the observed color (e.g., $\bv$) of a star with a
particular $Q$ value, and the intrinsic color (e.g., $(\bv)_0$), of an
unreddened star with the same $Q$ value (\cf Figure \ref{fig:qbv}).
The intrinsic sequences for OB stars in Figure \ref{fig:qbv} are
determined by fitting a line in each $Q$-color diagram to a
compilation of observations of unreddened Galactic OB stars
(\cite{str92}).  The {\it rms} dispersions of the Strai\v{z}ys data
about the best-fit lines are 0.006 mag and 0.004 mag for $Q_1$ and $Q_2$,
respectively.  The offset of the vast majority of stars in our sample 
toward redder colors indicates that most of these stars are reddened.
The correspondence between the LOSRV derived from the two $Q$'s is
illustrated in Figure \ref{fig:q1q2}.  We have removed stars from the
sample if their reddening values as derived from $Q_1$ and $Q_2$ differ
at more than the $5\sigma$ level, or if $E(\bv) < -5\sigma$, where
$\sigma=0.039$ mag, the mean propagated LOSRV error.  If a star is
rejected by either of these conditions, it suggests either that the
star is unusual or that there is a problem with the data for that
star.  Twenty-seven stars (out of 2619) were rejected for this reason.

The standard deviation of the remaining stars from the line
$E(\bv)_{Q_2}=E(\bv)_{Q_1}$ is 0.044 mag, only slightly larger than 
the mean propagated LOSRV error ($\sigma=0.039$ mag). The small
additional scatter introduced by the two different calculations of
$E(\bv)$ demonstrates that our choice of coefficients for $Q_1$ and
$Q_2$ is reasonable and that the extinction curve in the observed
region of the LMC is not grossly different from the adopted curve.
We adopt the mean of the two color excesses, weighted inversely by their
uncertainty, as the LOSRV in our subsequent analysis and refer to this
quantity as $E(\bv)$.  Because $Q_1$ and $Q_2$ are not entirely
independent, we adopt the smaller of the propagated uncertainties of
$E(\bv)_{Q_1}$ and $E(\bv)_{Q_2}$, rather than the uncertainty of the
mean, for the uncertainty in $E(\bv)$.

Finally, we de-redden the OB stellar photometry using the LOSRV and note
that most stars collapse tightly to the intrinsic main sequence
(\cf Figure \ref{fig:dred}), but that a surprisingly large fraction 
($\sim$ 20\%) collapse instead to $\bv = 0$ (open circles in Figure 
\ref{fig:dred}).  These objects can also be seen as a broad vertical 
band in Figure \ref{fig:qbv} at $Q\sim 0$.
Their coherence in Figure \ref{fig:qbv}, their uniform spatial 
distribution, their spread in magnitudes, and the consistent reddening
determinations from $Q_1$ and $Q_2$ all suggest that these objects are
not the result of faulty data or reduction problems.  Examination of
these stars in the two-color diagram shows
that they are probably supergiant stars in the LMC.  While these
supergiants are degenerate with reddened OB stars in the
color-magnitude diagram (CMD), the degeneracy is broken in the two-color
diagram.  The supergiants lie along a sequence in the two-color
diagram that is nearly parallel to the reddening line, causing the
de-reddening algorithm to collapse all of these stars to the same
$\bv$ color.  This behavior allows us to easily  
exclude these stars from our sample, by keeping only those stars with
$\bv<-0.1$ mag after de-reddening ($(\bv)_0=-0.17$ mag for spectral type B5,
the latest type allowed by our faint magnitude limit).  After this
final cut, we are left with 2069 OB main sequence stars with measured
LOSRV in this section of the LMC (from an initial sample of 2619
candidate OB stars). 

We test the validity of our estimated uncertainties by 
examining the distribution of the LOSRV (\cf Figure \ref{fig:hist}).
Had we severely underestimated the errors, a larger than expected
number of stars would have scattered toward unphysical, negative inferred
reddening values.  The stars with $E(\bv) < 0$ are well-fit by the
left half of a Gaussian centered at zero, with a FWHM characterized by
the propagated uncertainties ($\langle \sigma \rangle = 0.039$ mag),
as shown by the solid curve in Figure \ref{fig:hist}.
Therefore, they are consistent with a small, zero-reddening population
plus photometric errors.

Finally, because we are interested in the internal distribution of dust
in the LMC, we need to correct the reddening values for foreground
Galactic extinction.  We use the foreground reddening map for the LMC
of Oestreicher \etal (1995).  This foreground map has 10 arcmin
pixels and is based on the photometry of 1409 Galactic stars in the
direction of the LMC.  Figure \ref{fig:hist} also shows the
foreground-corrected histogram.  The inclusion of the Oestreicher
\etal uncertainties into our error budget increases the mean
propagated LOSRV error to $\langle \sigma \rangle = 0.045$ mag. 
Note in Figure \ref{fig:hist} that the stars with $E(\bv) < 0$ are
still fairly well-fit by a $\sigma=0.045$ Gaussian centered at zero, 
again consistent with the presence of a zero-reddening population plus
photometric errors.  The amplitude of this Gaussian is larger, simply
because there are now more stars with $E(\bv) < 0$ than there were
before the foreground extinction correction. 

\subsection{Highly Obscured Stars}\label{sec:obsc}

Our photometric selection criteria bias our sample against highly
obscured stars.  We originally excluded stars with $\bv > 0.4$ 
mag, because these are photometrically degenerate with a population of
Galactic main sequence stars, and so we cannot detect OB main sequence 
stars with $E(\bv) \gtorder 0.7$ mag.  

To determine whether there is a significant population of such stars,
we perform a ``pseudo-dereddening'' of the photometry of red giant
branch stars that is similar to blue envelope techniques used to
determine the reddening of  LMC supergiants.  We cannot perform a true
reddening analysis of these stars, because the intrinsic colors of
these red giant stars are not independently known; the blue envelope
is assumed to be the reddening zeropoint.  However, the assumption of
a small mean extinction is not likely to be wrong by more than 0.1 or
0.2 magnitudes, because the position of the red clump is approximately 
where it is expected, for $(m-M)_{LMC}=18.5$ mag.  We isolate a region
of the CMD bounded by two reddening lines, and by a line selected to
follow the blue envelope of the red giant branch.  This region is
shown in Figure \ref{fig:hess}.  We exclude stars with
large photometric errors ($> 0.2$ mag for $U$; $> 0.1$ mag for
$B,V,I$) from this analysis.  We then slide each star in this region
along a reddening vector in the $\bv,V$ CMD until it lies on the blue
envelope of the red giant branch.  Figure \ref{fig:rgb} shows a
histogram of the resulting reddening values.  We find very few highly
obscured ($E(\bv)>0.7$ mag) red giant branch stars, even though stars
reddened by as much as $E(\bv)=1.2$ mag are above our detection limit.
We conclude that there are no large areas of high obscuration in this
region of the LMC, unless they are extremely obscured ($E(\bv)>1.2$
mag). 

Before discussing the geometric implications of the LOSRV
distribution, we finish our discussion of the procedure by presenting
the construction of a two-dimensional reddening map for this region of
the LMC. 

\subsection{Interpolating the LOSRV}\label{sec:interp}

The mean projected density of OB stars in this region of the LMC, 
roughly 0.25 OB stars per square arcmin, enables us to construct a 
map with an average resolution of $\sim$ 10 arcmin.  The scientific 
value of such a map depends on whether the observed LOSRV
are spatially correlated, and are therefore sampling coherent dust
structure in the LMC.  Incoherent effects may dominate, making the map
meaningless.  One such effect arises because we have collapsed a
three-dimensional distribution of dust into a two-dimensional map.  The
different LOSRV may simply indicate the relative depth of the OB stars
along the line-of-sight within a uniform dust layer.  Another
incoherent effect may arise if there are large-amplitude reddening
variations on small scales across the plane of the sky.  Such variations
would be undersampled by our algorithm, leading to the detection of false
structure (\cf \S\ref{sec:test}).

Even with these potential difficulties, the construction and
analysis of the map will help determine the degree to which
these issues affect our final inferences regarding the
reddening across the LMC.  In \S\ref{sec:struct}, we conclude that 
the map contains valuable information about the dust distribution in
the LMC, even though it is somewhat affected by both line-of-sight
depth effects and unresolved small-scale variations.  

The initial step in constructing the map is the selection of an
appropriate grid resolution, which is complicated by the non-uniform
distribution of OB stars (\cf Figure \ref{fig:obdens}).  To take
advantage of the resolution provided by localized high concentrations
of OB stars, we map the reddening of the OB stars onto a grid with
spacings of 1.2 arcmin in right ascension and declination.  However,
to avoid being dominated by noise in the region where OB stars are
scarce, we adopt an interpolation method that incorporates an adaptive
effective resolution scale.  This approach reduces the correlation
between the map uncertainty and the local surface density of OB stars.

We use the following algorithm to generate the reddening map.  We
search for OB stars with measured LOSRV within 
a circular area of sky of radius 1.2 arcmin centered on each
gridpoint.  If we find 20 or more LOSRV within the search area, we
perform a least-squares plane fit to the reddening values.  If we find
fewer than 20, then the search radius is increased by 1.2 arcmin, and
we repeat the search.  We continue increasing the search radius until
we find at least 20 LOSRV.  When 20 values are found, a plane is fit
and all grid points within the final search area are assigned the
value of the best-fit plane at their location.  Because search areas
overlap, each gridpoint will be assigned multiple interpolated
reddening values.  We adopt the weighted mean of these values for the
reddening at that grid point, and use the dispersion of the values
about the mean as an estimate of the reddening map error.  The
resulting reddening map is presented in Figure \ref{fig:redmap}a.  A
map of the dispersion of the assigned interpolated reddening values of
each gridpoint is shown in Figure \ref{fig:redmap}b.  

Figure \ref{fig:redmap} includes foreground extinction; it should be
used to correct LMC photometry.  However, to examine the internal
structure of dust in the LMC, it is necessary to use the
foreground-corrected LOSRV (\cf Figure \ref{fig:hist}).  The
corrected reddening map is shown in Figure \ref{fig:cormap}a, and the
map of the dispersion is shown in Figure \ref{fig:cormap}b.

\subsection{Tests of the Mapping Procedure}\label{sec:test}

The variation of the LOSRV along different lines of sight far exceeds
what is expected from the observational uncertainties.  We discuss the
nature of the structure visible in Figures \ref{fig:redmap} and
\ref{fig:cormap} in \S\ref{sec:struct}.  For now we simply test 
whether our interpolated map is an accurate representation of the
variations in the LOSRV.  We do so by attempting to recover artificial
reddening maps with our least-squares plane fitting algorithm.  We
start by taking the coordinates of our 2069 OB stars and assigning to
them LOSRV drawn from two-dimensional sinusoidal surfaces that
represent the artificial maps.  Each simulated LOSRV is assigned a 
random photometric error consistent with the observational
uncertainties.  We vary the 
amplitude and spatial frequency of the surfaces to derive limits on
our ability to detect and resolve reddening variations.  Some
representative reddening maps and the corresponding recovery
by our interpolation routine are shown in Figure \ref{fig:hypo}.

First, we examine our sensitivity to variation on different spatial
scales.  For these tests, the amplitude of the reddening variations is
held constant at $E(\bv)_{max}=0.7$, consistent with the variations in
the observed LOSRV.  In relatively dense regions, such as in an OB
association, we can trace variations on scales $\gtorder$ 7 arcmin,
while in the sparsest regions we can only trace variations on scales
$\gtorder$ 15 arcmin.  Even when the local variations are unresolved,
the mean reddening value is recovered, except for the false structure
due to undersampling in the very high frequency map (\cf
\S\ref{sec:coh}).  If the actual reddening in low stellar density
regions varies with as much amplitude (0.7 mag) on such small scales
($\simless 7$ arcmin), then our map may contain false structure.  

Next, we examine our sensitivity to different amplitudes.  
The wavelength of the variations is held fixed at approximately 15
arcmin, and we examine the recovery of amplitudes set to $8\sigma$,
$4\sigma$, $2\sigma$ and $1\sigma$.  In each case, the surfaces are
normalized to the same mean reddening value, and the greyscale in
Figure \ref{fig:hypo} is held constant.  We find that the technique  
is sensitive to $> 2\sigma$ reddening amplitudes, where $\sigma$ is
the propagated LOSRV error.  Even when these low amplitude variations
are not recovered, the mean value of the reddening in that region is
recovered.  

Our final test of the interpolation technique, a self-consistency check,
is the comparison of the LOSRV of OB main sequence stars to the
reddening values estimated from the map at the position of the OB
stars (\cf Figure \ref{fig:qmap}). For stars with $E(\bv)_Q \simless
0.2$, the two determinations correlate, with a zeropoint offset of
0.021 mag.  The {\it rms} scatter of these low-reddening stars about
the best fit line (the dotted line in Figure \ref{fig:qmap}) is 0.056,
which is consistent with the characteristic error of the reddening map
values ($\sigma=0.06$).  Above $E(\bv)=0.2$, the interpolation
systematically underestimates the LOSRV because the algorithm smooths
over sharp peaks in the LOSRV distribution.  This smoothing effect is
less of a problem in high density, high LOSRV regions.  For example,
the peak in the reddening map (Figure \ref{fig:redmap}a) near
$(5.25^h,-67.4\degrees)$ corresponds to a dense OB association.  Most
of the high LOSRV stars lie outside of OB associations and are 
surrounded by stars with lower LOSRV, which may indicate highly
localized reddening ({\it e.g.}, circumstellar envelopes).  If these
high LOSRV are truly localized, the map is a more accurate
representation of the general reddening distribution than one might
infer from Figure \ref{fig:qmap}.

\section{Discussion}\label{sec:discuss}

We will use the distribution of LOSRV and the reddening map to perform
several investigations of the dust distribution in the LMC.  First, we
compare our LOSRV distribution to results from previous studies of 
reddening in the LMC (\S\ref{sec:comp}).  Second, we use our data to
place constraints on the line-of-sight geometry of stars and dust in
the LMC (\S\ref{sec:geom}).  Third, we test an analytic technique used
to correct the luminosity of distant galaxies for internal extinction
(\S\ref{sec:ext}).  Fourth, we investigate the nature of the observed
LOSRV variations and subsequent map structure (\S\ref{sec:struct}).
We examine the effects of line-of-sight depth differences and 
small-scale dust structure, the correlation of the LOSRV with OB
star density, and the spatial coherence of the LOSRV.  We conclude
that a positionally-dependent reddening correction does provide some
improvement to the single-value reddening correction and that it can
highlight regions of potentially high extinction, as well as regions of
favorable low extinction.  We have not, and cannot, fully account for 
the line-of-sight depth effects, the undersampling effects, or the
possible bias introduced by sampling the ISM with massive stars.
Nevertheless, the LOSRV and the reddening map provide the most
detailed information on the distribution of dust in this region of the
LMC yet available.  Finally, we apply a positionally-dependent
reddening correction to the photometry of this region
(\S\ref{sec:nonob}). 

\subsection{Comparison to Previous Studies}\label{sec:comp}

Previous studies have typically found that the  mean reddening toward
LMC stars is $\ltorder$ 0.1 mag (\cf \cite{fea60}, \cite{iss75},
\cite{gm86}).  More recently, Hill \etal (1994) found $\langle
E(\bv)\rangle =0.15$ mag, Massey \etal (1995) found $\langle
E(\bv)\rangle =0.13$ mag and Oestreicher and Schmidt-Kaler (1996)
found $\langle E(\bv) \rangle =0.16$ mag, although Hill \etal focused
on OB associations, rather than the entirety of the LMC.  It should
also be noted that the Hill \etal and Massey \etal values are not
corrected for foreground extinction, while the Oestreicher and
Schmidt-Kaler value is corrected.  Oestreicher 
and Schmidt-Kaler argue that other studies underestimate the mean
reddening either because they use Galactic intrinsic stellar colors
for supergiants, or because the faintest stars are biased toward low
reddening values because of a color-independent magnitude limit.  The
former causes a bias because supergiants in the LMC are intrinsically
bluer than their Galactic counterparts due to their lower metallicity
(\cite{osk96}).  The latter causes a bias because highly reddened,
faint stars are preferentially lost below the faint magnitude limit.
We avoid the supergiant color issue by only using main sequence OB
stars, whose intrinsic colors are insensitive to metallicity changes
(\cite{ogs95}).  We address the completeness issue by introducing a 
color-dependent limiting magnitude cut that is not biased against
faint, reddened stars relative to faint, unreddened stars.  We are
able to construct this color-dependent cut because the faintest OB
stars are many magnitudes brighter than our limiting magnitude.  
Our analysis (corrected for foreground Galactic extinction)  yields
$\langle E(\bv) \rangle =0.13$ mag; consistent with the most recent
results.  However, our mean value is based on only $\sim$ 4 percent of
our eventual total LMC survey area, and this area may be
systematically different from the majority of the LMC. 

In addition to presenting the mean value, two of the recent studies
have included 
histograms of the LOSRV.  The position and width of our low-reddening
peak is consistent with the LOSRV distributions observed by
Oestreicher and Schmidt-Kaler and Massey \etal  However, Massey \etal 
do not find a significant high-reddening tail.  This may be partly
due to their smaller sample size, but when we renormalize our
histogram to match their number of OB stars, we still find a
more significant tail in our data.  We suspect that the discrepancy exists 
because the Massey \etal  sample does not employ a color-dependent
faint magnitude limit; thus, it is biased against the detection of
highly reddened stars near the faint magnitude limit.  Furthermore,
they require spectroscopy to determine the intrinsic colors, so their
limiting magnitude is quite bright (roughly $V=15$ mag).  When we
limit our sample with a color-independent magnitude limit at $V=15$
mag, we obtain a similar histogram to that of Massey \etal  

\subsection{Dust Geometry Models}\label{sec:geom}

The LOSRV are a measurement of the total column density of dust 
toward the OB stars. The LOSRV depend both on the
physical density of the ISM, and the line-of-sight depth of
the star inside the dust layer.  In this section, we discuss our 
constraints on the relative distribution of stars and dust along the
line-of-sight in the LMC by comparing the observed LOSRV distribution
(Figure \ref{fig:hist}) to those produced by models of various star
and dust distributions.  We begin with simple models in which the
stars are uniformly distributed in a slab of finite thickness, and the
dust is either in a slab that lies in front of the stars, uniformly
mixed with the stars, or in an infinitesimally thin midplane sheet. We
also examine the more realistic situation of exponential disk
distributions of stars and 
dust.  We allow the scale height of the stars and dust to vary
independently in these models, and present three representative cases
here:  ${h_s}/{h_d}$ = 0.5, 1.0, and 2.0, where $h_s$ is the scale
height of stars and $h_d$ is the scale height of dust.  In each of
these six models, we generate the intrinsic stellar photometry from a
theoretical isochrone (\cite{bbc94}; age=4 Myr (their youngest)), add
photometric errors, place the star within the geometrical model, and
redden its colors in proportion to the amount of foreground dust
according to the LMC extinction curve.  We define the mean optical
depth in each model to match the observed mean ($E(\bv)=0.13$ mag).
The simulated LOSRV distributions are shown in Figure \ref{fig:geom}.

The slab models produce LOSRV distributions markedly different from
that observed.  Aside from the unphysical premise of foreground sheet
model, the resulting LOSRV distribution is much narrower than the
observed distribution.  This result is non-trivial because standard
mean-value reddening corrections to LMC photometry imply this
geometry.  The midplane sheet model generates a wider LOSRV
distribution, but the distribution is strongly bimodal $-$ half of the
stars are reddened a non-zero constant amount, and the other half are
unreddened. Of the three slab models, the mixed model produces the
best match to the observations, but the asymmetric tail to high LOSRV
is still not reproduced and the peak in the distribution is too
flat. A more complex model is needed. 

Because the LMC is a disk galaxy (\cf \cite{pre89} and references
therein), we expect the stars and dust to be well approximated by a
vertical exponential distribution (\cite{bs65}, \cite{dg96}).  We
model the distribution of stars and dust along the line-of-sight as
exponentials about a fiducial midplane that is perpendicular to the
line-of-sight.  The inclination of the LMC is between $\sim
33\degrees$ and $45\degrees$ (\cite{wes90}), but the inclination only
introduces a constant factor of sin($i$) to both the star and dust
scale heights.  Since the ratio of the scale 
heights of the stellar and dust distributions is the only free
parameter of the model, the inclination does not affect our analysis.
As ${h_s}/{h_d} \rightarrow 0$, the model reduces to the foreground
sheet model (since all the stars are equally reddened); as
${h_s}/{h_d} \rightarrow \infty$, it reduces to the thin midplane
layer model; and intermediate models resemble the mixed model.  As one
increases $h_s$ relative to $h_d$ the reddening distribution will
widen, but it will also become increasingly bimodal. Again, we set the
mean of the model LOSRV distribution to match the data ($\langle
E(\bv)\rangle =0.13$).  Finally, we stress that $h_s$ refers to the
scale height of OB stars, not the general population of LMC stars. 

Many dust models invoke a plane-parallel geometry (\cf \cite{brz88},
\cite{dib95}) or an exponential disk geometry (\cf \cite{dis89},
\cite{byu94}, \cite{cor96}).  However, one geometrical aspect that poses 
serious difficulties to both empirical and theoretical treatments of
this problem, and has therefore generally been ignored, is clumpiness
in the distribution of stars and/or dust (for exceptions, see
\cite{boi90}, \cite{hs93}).  As illustrated in Figure \ref{fig:geom}, 
the model with $h_s/h_d = 0.5$ best matches the width of the observed
distribution peak.  This ratio is consistent with the vertical
structure of most spiral galaxies.  Allen (1973) found that the
typical OB stars-to-dust scale height ratio is 0.42.  However,
this smooth exponential model fails to account for both the tail of
highly reddened stars and the number of stars with $E(\bv)<0$ mag.
The simplest modification of the model is to invoke a clumpy component
to the distribution of dust. 

To simulate a clumpy dust distribution, we randomly select a fraction
of the lines-of-sight to be more highly reddened than the exponential
dust envelope would predict.  These lines of sight are given LOSRV
drawn from either (a) an exponential or (b) an offset Gaussian
reddening distribution.  We also add ``holes'' to the distribution to
match the data with $E(\bv)<0$.  The holes are modeled simply by 
randomly selecting a fraction of the lines-of-sight to have zero
reddening.  These clumpy-model parameters are interactively adjusted
until the models match the data.  In the exponential model,
40\% of the stars are assigned the midplane reddening, plus an
additional reddening drawn randomly from an exponential
distribution with a reddening scale length of 0.17 mag.  
In the Gaussian model, 50\% of the stars have reddening drawn from a
Gaussian with a mean value of $0.2$ mag and $\sigma=0.2$ mag (the
model excludes Gaussian reddening values below zero).  In both cases,
we set 5\% of the lines-of-sight to have zero reddening.  
There is yet no physical motivation for these particular model
parameters.  The requirement of a clumpy component 
suggests that at least some of the structure in the reddening map is
real, and that approximately half of the OB stars have reddening
values that cannot be explained by a smooth exponential disk model.

\subsection{Internal Extinction in Spiral Galaxies}\label{sec:ext}

The LOSRV and the geometric models of dust and star distributions
allow us to derive the total optical depth through the LMC and to
compare the extinction as derived on a star-by-star basis 
to that derived from global properties.  This provides an independent
check on whether the integrated magnitudes of more distant galaxies
are being properly corrected for internal extinction.

We now derive a quantitative measurement of the total $B$-band optical 
depth for the LMC, which we present as new evidence for the ongoing
debate over the optical thickness of disk galaxies (see \S1). 
We begin with our midplane
reddening measurement, $E(\bv)=0.13$ mag. By applying the standard
Galactic optical extinction curve (fine for the LMC (\cite{fit85})),
this reddening is equivalent to a $B$-band extinction, $A_B$, of 0.533
mag.  This extinction is directly transformed to the optical depth
using $\tau_B=\frac{A_B}{1.086}=0.49$.  The total optical depth of the
observed region of the LMC disk is assumed to be twice this midplane
value.  The face-on optical depth is given by $\tau_B^0 =0.98 \times
cos(i)$.  The inclination of the LMC is between $33\degrees$ and
$45\degrees$, therefore the face-on optical depth of the LMC disk is
between $\tau_B^0=0.82$ and $\tau_B^0=0.69$. 

This optical depth measurement can be compared to an extinction
correction for spiral disks provided by Tully and Fouqu\'{e} (1985,
hereafter TF). There may be concerns about applying such a correction
to an irregular galaxy like the LMC; however TF themselves include
such galaxies in their sample.  Furthermore, the only property 
required of a galaxy for this model to work is that it be a disk
system. The extinction correction is based on a finite, plane-parallel 
midplane absorbing layer, and it is given by: 

$$A_B = -2.5 \log [f(1 + e^{-\tau_B \sec(i)}) + (1 - 2f)(\frac{1-e^{-\tau_B \sec(i)}}{\tau_B sec(i)})],$$

where $\tau_B$ is the face-on optical depth of the absorbing layer, and
$(1-2f)$ is the fractional thickness of this layer, compared to the
thickness of the stellar slab (for simplicity, stars of all types have
the same vertical distribution).  TF adopt $\tau_B=0.55$ and $f=0.25$
as the best-fit values to their sample of 600 nearby spiral galaxies,
independent of galaxy type.  These parameters are poorly constrained.
According to their Figure 5, values of $\tau_B$ between 0.4 and 1.2,
and values of $f$ between 0.0 and 0.4 are acceptable.  In principle,
this equation could be used to derive a galaxy's true, extinction-free
luminosity.  In practice, because of the poor constraint on $\tau_B$,
it is used only to correct a galaxy's photometry to its face-on
values.  The correction from face-on magnitudes to dust-free
magnitudes has remained largely unknown.  

While our optical depth measurement is higher than TF's prediction,
it is well within their acceptable range.  A straightforward
application of the mean TF correction to the integrated photometry of
the LMC would result in a
16\% underestimate of the $B$-band luminosity of the LMC, compared
with a correction based on our $B$-band extinction. Recently, Tully
\etal (1997) have refined their model parameters to values of
$\tau_B=0.8$ and $f=0.1$, and find a dependence of these values on
galaxy type.  Their new calibration would suggest that a 
galaxy like the LMC should have $\tau_B \sim 0.4$, resulting in a 22\%
underestimate of the $B$ luminosity.  We conclude that ({\bf 1}) the
optical depth through the LMC is relatively high ($\tau_B \sim 0.75$)
even at a radius of 2 kpc and disregarding the tail of high LOSRV
(this is independent of the validity of the TF model), ({\bf 2}) that
if SBm galaxies do indeed have lower optical depths than larger
spirals (as suggested by \cite{tul97}), then the latter likely have
$\tau_B \sim 1$, and ({\bf 3}) that the statistical reddening correction
presented by TF is valid, at least for this region of the LMC, to
$\sim 20\%$. 

These results come with two important caveats.  First, our optical
depth measurement is based on the reddening of OB stars.  This
population may be in regions of higher than average extinction, in
which case we have overestimated the mean optical depth.  Second, 
investigators generally use a correction like that in TF to correct 
galaxies to face-on magnitudes, rather than to dust-free magnitudes, 
and relative reddening differences may be more precise than the $20\%$
uncertainty found above.

\subsection{The Nature of Observed Reddening Variations}\label{sec:struct}

To determine whether the reddening map is useful for de-reddening
observations within this region, we need to establish
whether incoherent effects, such as different line-of-sight depths of
the OB stars and small-scale dust density variations, dominate the
observed structure.  Several lines of argument will lead to the
same conclusion: incoherent effects of line-of-sight depth
differences and small-scale reddening variations are present, but they
do not dominate the observed structure in the reddening map.

\subsubsection{Line-Of-Sight Depth Effects}\label{sec:losde}

First, we must determine how much of the structure seen in the
reddening map is due to line-of-sight depth differences among the OB
stars, rather than to density variations in the dust across the
line-of-sight.  If the observed structure is dominated by
line-of-sight-depth effects, we would expect the following: (1) the
line-of-sight models constructed in \S\ref{sec:geom} should reproduce
the observed LOSRV distribution without a need to invoke clumpiness; 
(2) OB stars in associations should be similarly
reddened, while field OB stars should have higher LOSRV dispersion;
(3) the LOSRV of each OB star should indicate its relative depth along
the line-of-sight; and
(4) random re-assignment of the LOSRV should not affect the spatial
correlation of the LOSRV.  We have already demonstrated
(\S\ref{sec:geom}) that line-of-sight depth differences in a
smooth dust envelope cannot reproduce the observed LOSRV histogram.  
We address (2) and (3) next, and (4) in \S\ref{sec:coh}.

OB stars in associations will be similarly reddened if line-of-sight
depth effects dominate the LOSRV, since the OB stars in an association
are localized along the line-of-sight.  Field OB stars are, in
general, distributed widely along the line-of-sight.  They will
therefore have larger LOSRV dispersion.  The two obvious OB
associations in this area of the LMC have differential reddening that
is at least as large as that among the field OB stars (\cf Figure
\ref{fig:depth}).

If depth effects dominate the LOSRV, then Figure
\ref{fig:depth} can be interpreted as a plot of the spatial
distribution of OB stars in the LMC along the line-of-sight.  The
unphysical, apparently larger, extent of the far side of the LMC
implied by this Figure is an indication that, at least for the high
LOSRV, line-of-sight depth differences cannot fully explain the
observed reddening, even for the field OB stars.

\subsubsection{The Reddening Coherence Length}\label{sec:coh}

As a final test of the clumpiness in the dust distribution, we measure 
the coherence length of the LOSRV.  Small scale variations in the dust
distribution can produce false structure in our reddening map (\cf
Figure \ref{fig:hypo}).  The LOSRV will not be spatially coherent if
depth differences or these small-scale variations dominate.  We
attempt to measure a reddening coherence length by calculating a
correlation function for the LOSRV in the following manner.  Beginning
with the LOSRV for one star, $L_1$, we calculate $|L_1 - L_i|$, where
$L_i$ represents the reddening for the $i$th star.  We do this for
every star, considering each pair once.  The pairs are binned by their
angular separation, and the median reddening difference among the
pairs in each bin is taken as characteristic for that separation.  
This procedure is carried out for four populations of OB stars:  all
OB stars, field OB stars only, low-reddening OB stars, and
low-reddening field OB stars.  These populations allow us to
determine if the coherent structure is restricted to high reddening
values, or to areas of high stellar density (OB associations).  We
also randomly re-assign the LOSRV in each of these four populations
and calculate the median LOSRV difference at each radius.  The
correlation is then characterized by the ratio of the median
difference in the original data to that in the randomized data.
The results are shown in Figure \ref{fig:coh}.  

The decline in the normalized absolute difference as $r \rightarrow 0$ 
indicates that the LOSRV are correlated at small separations.  The
lack of a perfect correlation as $r \rightarrow 0$ ({\it i.e.} the
normalized difference does not go to zero) indicates that either
line-of-sight depth effects, small-scale variations, or both are
present in the dust distribution.  Figure \ref{fig:coh} also
demonstrates that there is little or no coherence beyond $\sim 20$
arcmin (the planes used to construct our reddening map are always fit
over a region smaller than 20 arcmin), suggesting that maps with
resolution $\gtorder 20$ arcmin contain little spatial reddening
information. 
The anti-correlation beyond one degree of separation may indicate a
global reddening gradient in this field.  Again, we conclude that the
observed structure in the reddening map in part represents true
structure in the dust distribution.  All four populations of OB stars
exhibit similar coherence properties, indicating that coherent
reddening structure can be inferred, even among the field population
of OB stars, and even when the LOSRV are rather low ($E(\bv) < 0.2$
mag).  

\subsubsection{Correlation of the Reddening with OB Star
Density}\label{sec:corl} 

Because the dust distribution is somewhat clumped, one might
expect that areas with a locally high star formation rate, which
correlate spatially with dense areas of the ISM (\cite{sch59}), 
might lie in regions of higher than average dust density.  We find
that there is no significant correlation between LOSRV and projected
OB star density, except possibly at low stellar densities.
Likewise, previous studies of reddening in OB associations (e.g.,
\cite{hmf94}) have not found higher average reddening than studies of
field stars (e.g., \cite{osk96}, \cite{mass95}).  A comparison of Figures
\ref{fig:obdens} and \ref{fig:redmap} reveals that although regions of
high OB star density often have relatively low reddening, they tend to
have nearby regions of high reddening.  This may be an indication of
the early stages of the disruption of the ISM caused by the
high-energy radiation of OB stars.  In particular, if we construct a
high-resolution reddening map for the region surrounding the largest OB
association (near $\alpha=5.23^h, \delta=-67.4\degrees$), we see this
effect in detail.  We improved the resolution by requiring six stars
instead of twenty for each plane fit, and increasing the density of
gridpoints by a factor of four.  This map is shown in Figure
\ref{fig:himap}, and the approximate position of the center of the OB
association is marked.  It shows a shell-like structure of high
reddening just south of the center of the OB association, with
relatively low reddening at the association's center.

\subsection{The Reddening Correction for Non-OB Stars}\label{sec:nonob}

Since the reddening map is, at least in part, tracing coherent dust
structure in the LMC, 
we use the map to de-redden all 1.1 million stars in this section of
the LMC.  To estimate the reddening of each star, we perform a
bi-linear interpolation of the four nearest map pixel values. As a
test of how well our positionally-dependent correction worked, we
reconstruct the $BV$ CMD and examine the width of the de-reddened
stellar sequences.  We find that the sequences are generally about the
same width as before; the upper main sequence has tightened up, but
the red giant branch has gotten slightly wider.  This indicates that
the correction estimates have not been dominated by line-of-sight
depth effects or small-scale variations, but also that the map is not
a full description of the dust distribution.  In particular, the
reddening for older stars, {\it e.g.} red giants, appears to be
slightly different than that of the OB stars.  This may be an
indication of the migration of these older stars away from the dense
regions of the ISM.

\section{Summary}\label{sec:summ}

Based on the $UBVI$ photometry of main sequence OB stars, we
have measured the line-of-sight reddening along 2069 lines-of-sight in
the LMC.  The distribution of these reddening values reveals that
there is significant differential reddening in the LMC.  We find that
this can be partially explained by the different line-of-sight depths
of the OB stars inside a smooth envelope of dust, but that this
explanation falls short of fully accounting for the shape and width of
the reddening distribution.  We conclude that there is some structure
in the distribution of the dust.  Although this may seem trivial 
given our knowledge of dust in our Galaxy, all current reddening
corrections and most models of external galaxies adopt a smooth dust
distribution.  The LMC provides an opportunity to test these models.

We constructed line-of-sight geometric models to constrain the relative
distributions of stars and dust perpendicular to the LMC midplane.  We
found that the best model has exponentially distributed stars and dust,
with the dust scale height twice that of the OB stars.  We also found
that the tail to high line-of-sight reddening values (LOSRV) could not
be reproduced by any simple line-of-sight geometry.  The optical depth
through this region of the LMC is $0.69 < \tau_B < 0.82$, for an
inclination angle $33\degrees < i < 45\degrees$.

We used our measurement of optical depth of the LMC to investigate how
well a straightforward application of a semi-empirical extinction
correction like that provided by Tully and Fouqu\'e 1985 (TF) recovers
the dust-free luminosity of this galaxy.  We found that the LMC
contains more obscuring material than is predicted by this correction,
although the optical depth estimates of TF are quite uncertain and the
optical depth inferred from OB stars may be an overestimate.

We examined whether the structure in the reddening map
is due to coherent reddening variations across the LMC's projected
surface, or to line-of-sight depth effects and unresolved small-scale 
variations in the dust density.  We found that the latter effects are not
entirely responsible for the observed structure for four reasons. 
First, in a smooth-envelope model where the observed stucture is due
to line-of-sight depth differences, stars in associations should be
more uniformly reddened than stars in the field.  We did not observe
this effect.  Second, if the reddening is proportional to the 
star's position along the line-of-sight within the dust envelope, then
the reddening can be used to infer each star's depth in the LMC.
However, the inferred line-of-sight  
distribution has unphysical structure.  Third, our dust-star geometric
models are only partially successful in reproducing the observed LOSRV
distribution.  In particular, fitting the tail of high LOSRV required
a clumpy component of dust.  Fourth, we found that the LOSRV are
partially coherent for angular separations $\ltorder$ 20 arcminutes.  
This result indicates that the observed reddening structure is not
dominated by either line-of sight effects or unresolved variations,
both of which should produce incoherent LOSRV, at least among the
field OB stars.  

This work comes in a long line of studies of reddening in the
Magellanic Clouds based on observation of the most luminous member
stars.  We have been able to extend those studies by examining the
distribution of dust at high spatial resolution over a large region.  
It is evident that simple models for the distribution of dust fail to
match the observations.  There is, therefore, the potential for
misinterpretation of the integrated photometry of more distant
galaxies.  An understanding of the rich and complex structure of dust
in the Clouds is complicated by the unknown relative distributions of
stars and dust along the line-of-sight.  The combination of our full
survey (which will have approximately 20 times as much data as
presented here), infrared observations from ISO and IRAS, and HST
observations of galaxies seen through the Clouds should enable 
further progress. 

\vskip 0.7in
\noindent
Acknowledgments:
We gratefully acknowledge financial support from a NASA LTSA grant
(NAG-5-3501) and an NSF grant (AST-9619576), support for the
construction of the GCC from the Dudley Observatory through a Fullam
award and a seed grant from the Univ. of California for support during
the inception of this project.  We also thank the Carnegie Institution
for providing telescope access, shop time, and other support for this
project, and the staff of the Las Campanas observatory, in particular
Oscar Duhalde, for their usual excellent assistance.  Finally, we
thank M. O. Oestreicher for generously providing us with his
foreground reddening map data, R. Brent Tully for an extremely useful
conversation on the opacity of spiral disks, Katherine Wu for a
very helpful reading of an earlier draft of this work, and the
anonymous referee for careful and valuable comments.   

\vfill\eject

\vfill\eject
\centerline {\bf Figure Captions}
\bigskip

\figcaption{A $\bv$,$V$ Hess diagram from our LMC drift scan survey
image.  Over 1.1 million stars are represented.  Each star is
represented by a two-dimensional Gaussian with dimensions given by the 
observational uncertainties.  The vertical sequence labeled {\bf MS}
is the main sequence.  The area labelled {\bf RC} is the red clump.
The sequence labeled {\bf RG} is the red giant branch.  The vertical
sequence labeled {\bf FG} is a foreground population of Galactic stars
(\cf \S\ref{sec:select}).  The box in the upper left represents the
photometric limits used to isolate candidate OB stars (\cf
\S\ref{sec:select}).  The box in the lower right represents the
photometric limits used to isolate red giant branch stars (\cf
\S\ref{sec:obsc}).  The greyscale is logarithmic to emphasize low
surface brightness features. \label{fig:hess}}

\figcaption{Q-color plots for candidate OB stars in our
sample.  The left panel plots the Johnson and Morgan $Q_1$
parameter, adapted to the extinction curve of the LMC.  The right
panel plots our $Q_2$ parameter.  The lines are fits to the photometry
of unreddened Galactic OB stars.  The color excess, $E(\bv)$, of any
star is the vertical distance between that star and the
line. \label{fig:qbv}} 

\figcaption{The correlation of $E(\bv)$ as determined by the
Johnson and Morgan $Q_1$ parameter with $E(\bv)$ as determined by
our $Q_2$ parameter.  The line represents $E(\bv)_{Q_1}=E(\bv)_{Q_2}$.
The {\it rms} scatter about this line is 0.042 mag, while the
mean propagated error on our reddening measurements is 0.039 mag. 
\label{fig:q1q2}}

\figcaption{{\bf Left:} The uncorrected color-magnitude diagram for
our candidate OB stars.  {\bf Right:} The same CMD, but corrected for
interstellar reddening as outlined in the text.  The open circles
represent a population of stars that are apparently not OB main
sequence stars.  These stars are also visible in Figure \ref{fig:qbv}
as those with $Q > 0$. \label{fig:dred}} 

\figcaption{{\bf Left:} A histogram of $E(\bv)$ of OB stars as
determined from the $Q$ parameters.  The curve is a Gaussian
representing a population of zero-reddening stars with the same
propagated photometric errors as our data.  The curve has been
normalized to twice the number of observed OB stars with measured
reddening values less than zero. 
{\bf Right:} The histogram of $E(\bv)$, after correcting for
foreground extinction as determined by Oestreicher \etal (1995).  The
curve is again the prediction of negative reddening values based on
the propagated errors.  There is now a slight excess of stars with
negative LOSRV.  This may be due to unresolved clumpiness in the
foreground dust distribution. \label{fig:hist}}
  
\figcaption{The distribution of pseudo-LOSRV of red giant branch
stars.  The lack of highly-reddened stars is an indication that our
photometric selection criterion for OB stars ($\bv < 0.4$ mag) is not
biasing our results.  These values of $E(\bv)$ are relative to the blue
envelope of the red giant branch.  The true zeropoint of this
distribution is undetermined. \label{fig:rgb}}

\figcaption{The positions on the sky of the 2069 OB stars in our sample.
We attempt an interpolation of the reddening derived from these stars
to all other stars in the frame.  
\label{fig:obdens}}

\figcaption{{\bf (a)} The interpolated reddening map of our LMC scan.
We use this map to correct the photometry of non-OB stars in the
image.  {\bf (b)} The map of the dispersion of assigned planar
reddening values at each grid point.  In this and the following
figure, the greyscale is indicated by the vertical bar on the
right. \label{fig:redmap}} 

\figcaption{{\bf (a)} The interpolated reddening map, based on the
foreground-corrected LOSRV.  This map is a better indication of the
internal structure of dust in the LMC. {\bf (b)} The map of the
dispersion of planar reddening values, for the foreground-corrected
map. \label{fig:cormap}}

\figcaption{{\bf Top Row:} Four simulated dust distributions.  The
simulations show reddening variations with spatial wavelengths of,
from left to right: 240 arcminutes, 60 arcminutes, 15 arcminutes, and
7.5 arcminutes.  In each case, the amplitude of the reddening variations
is 0.7 magnitudes.  {\bf Center Row:} The recovery of these
simulations by our reddening interpolation algorithm.  Comparison to
Figure \ref{fig:obdens} illustrates that our resolution of small-scale
structure is highly dependent on the surface density of OB stars.  In 
the dense regions, we can resolve variations to better than 7.5
arcminutes.  {\bf Bottom Row:} The recovery by our interpolation
routine of the 15 arcminute simulation, with amplitudes of, from left
to right: 0.32, 0.16, 0.08, and 0.04 mag (the smallest amplitude
corresponds to $1\sigma$).  In each of these models, the pixel scale
has been normalized to match the amplitude of the 0.32 mag model.  
Our amplitude limit ($\sim$0.08 mag at these spatial wavelengths) 
is not highly dependent on the surface density of OB stars.  
In each of these simulations, the axes are the same as those in Figures 
\ref{fig:obdens}, \ref{fig:redmap}, and \ref{fig:cormap}. \label{fig:hypo}}  

\figcaption{Comparison of the reddening of OB stars as determined by
the $Q$ parameter analysis with the reddening of these stars as
re-calculated by the interpolation routine.   High LOSRV are not
reproduced in the map because they are generally uncorrelated.  
\label{fig:qmap}}

\figcaption{Line-of-sight geometry models. {\bf Top row:}  Plane-parallel
models.  Uniform foreground sheet (left), dust and stars uniformly
mixed (center), and infinitesmally thin midplane dust sheet (right).
{\bf Center row:} Exponential disk models.  $h_s/h_d = 0.5, 1.0, 2.0$
(left to right).  {\bf Bottom Row:}  Exponential disk, plus clumpy
component.  Exponential tail distribution (left) and secondary
gaussian distribution (center).  In the bottom right, we reproduce 
the foreground-corrected LOSRV histogram, for reference.  \label{fig:geom}}

\figcaption{LOSRV vs. Right Ascension for a slice of declination:
$-67.5\degrees < \delta < -67.2\degrees$.  If the dust layer is smooth,
then the reddening should indicate the relative depth of each star
along the line-of-sight.  The implied large extent of the far side of
the LMC and the elongation of the OB associations (indicated with
arrows) along the line-of-sight indicate that the dust layer must be
clumpy. \label{fig:depth}} 

\figcaption{The median difference between LOSRV pairs as a function of 
angular separation for:  all OB stars (upper left), field OB stars
(upper right), all low reddening OB stars (lower left), and low
reddening field OB stars (lower right).  In each case, the solid line
is the correlation of our measured LOSRV, and the dashed line is the 
correlation of the population after randomly reshuffling the LOSRV.
The differences are normalized such that the mean difference of the
randomized population is unity.  The plot indicates that there is
coherence in the reddening values out to separations of 20 arcminutes.
The anti-correlation at large separations may be due to a global
gradient in the LOSRV across the field.  \label{fig:coh}} 

\figcaption{A high-resolution reddening map of the largest OB
association in this section of the LMC.  The higher resolution was
achieved by using a higher density of map grid points, and by
requiring six (instead of 20) LOSRV for each plane fit.  These
changes are justified by the high density of OB stars in this area.
The cross marks the geometric centroid of the OB stars in this area
($5.21^h<\alpha<5.24^h$,$-67.5\degrees<\delta<-67.3\degrees$). Note
that this does not correspond to the area of highest reddening, and
that a high-reddening shell-like structure borders the
area. \label{fig:himap}}  

\vfill\eject

\begin{figure}[t]
\plotone{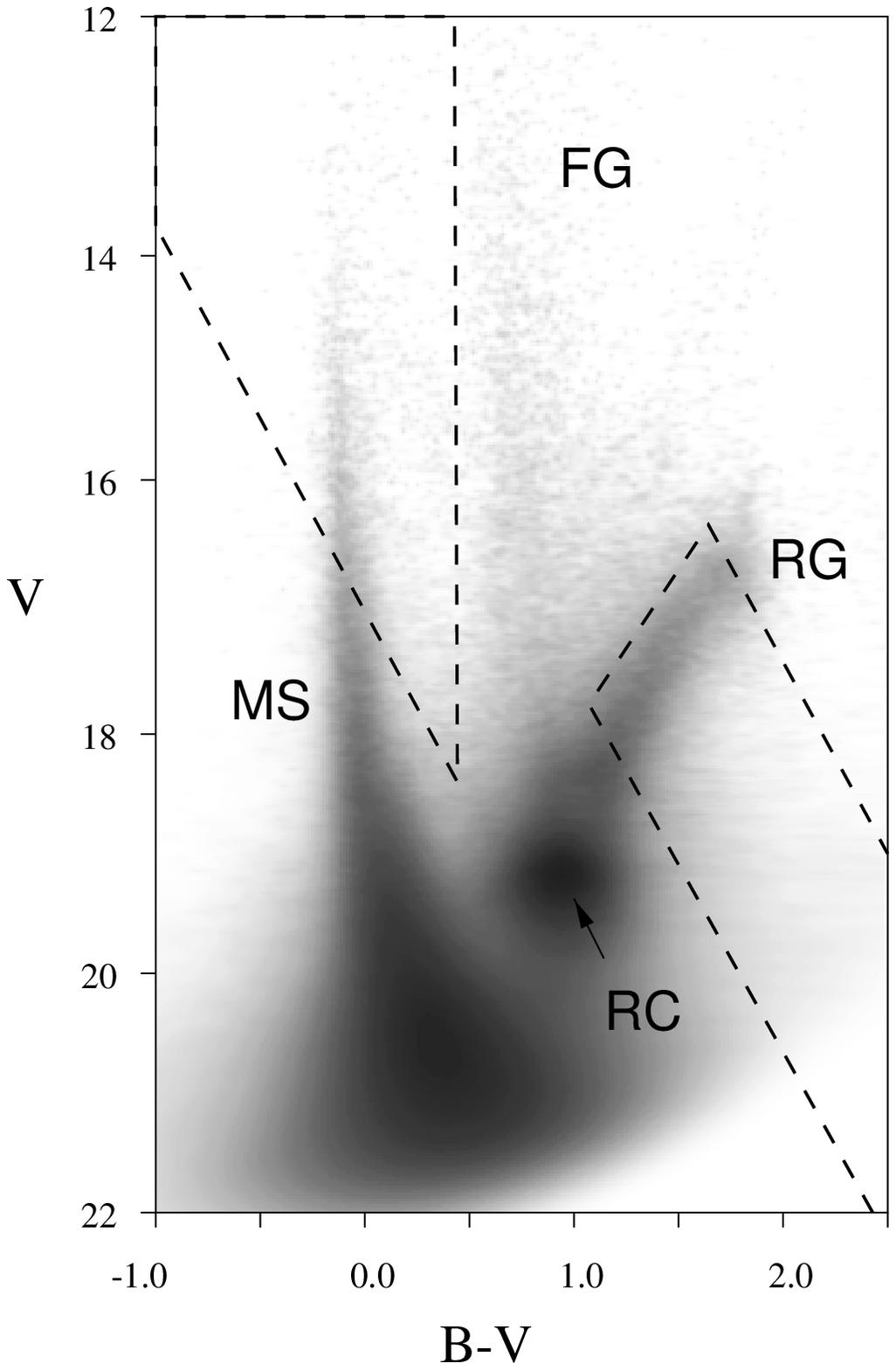}
\end{figure}

\begin{figure}[t]
\plotone{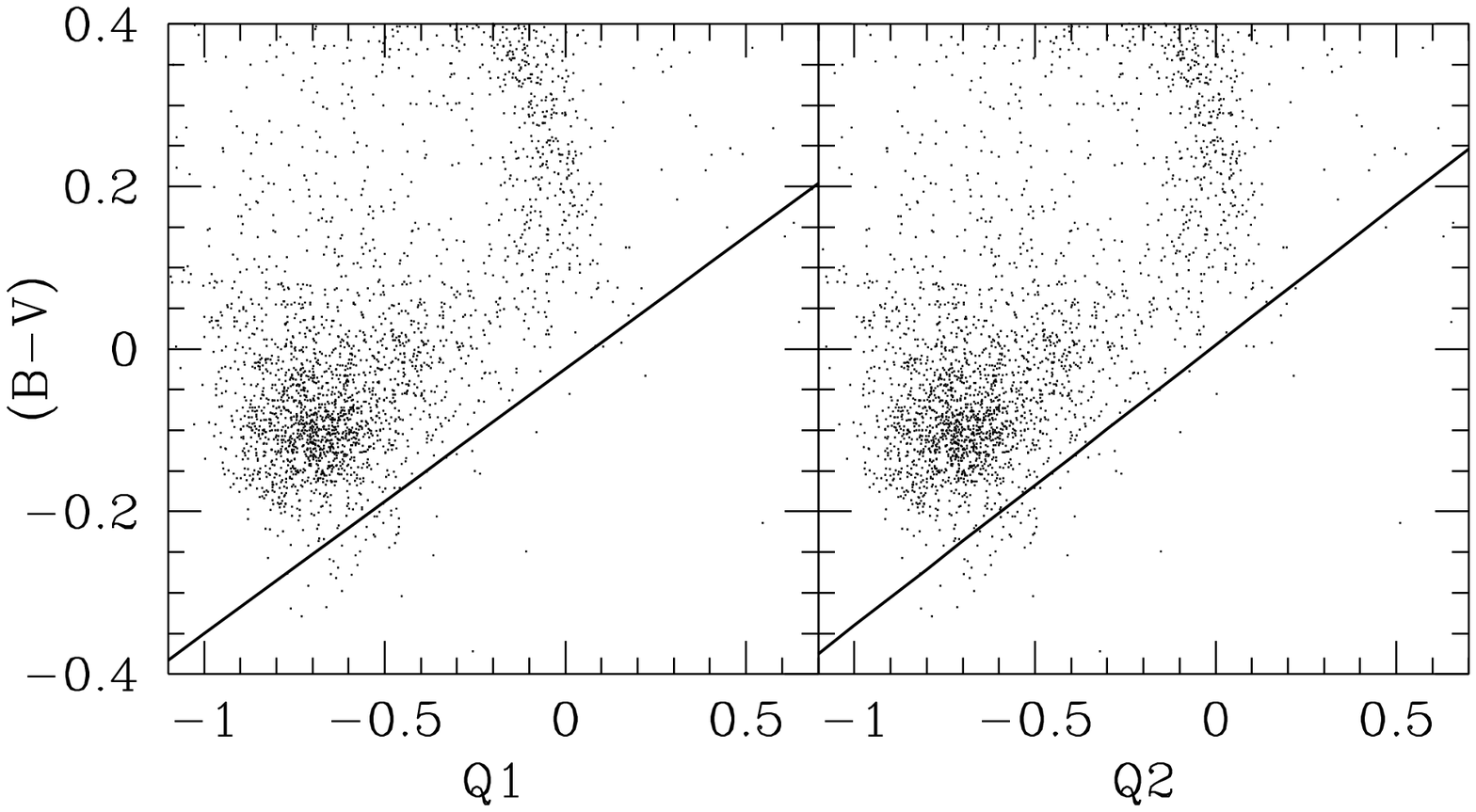}
\end{figure}

\begin{figure}[t]
\plotone{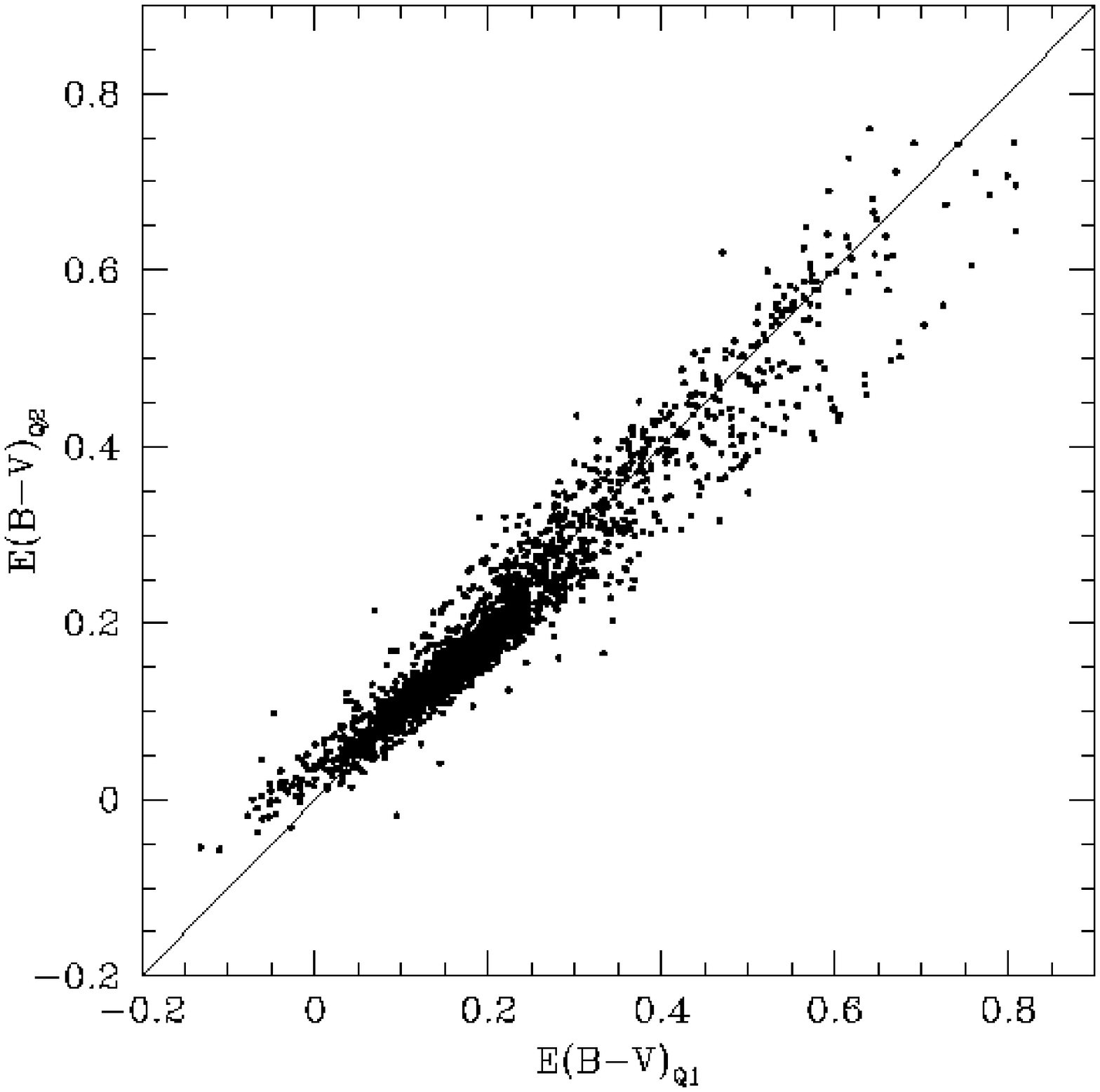}
\end{figure}

\begin{figure}[t]
\plotone{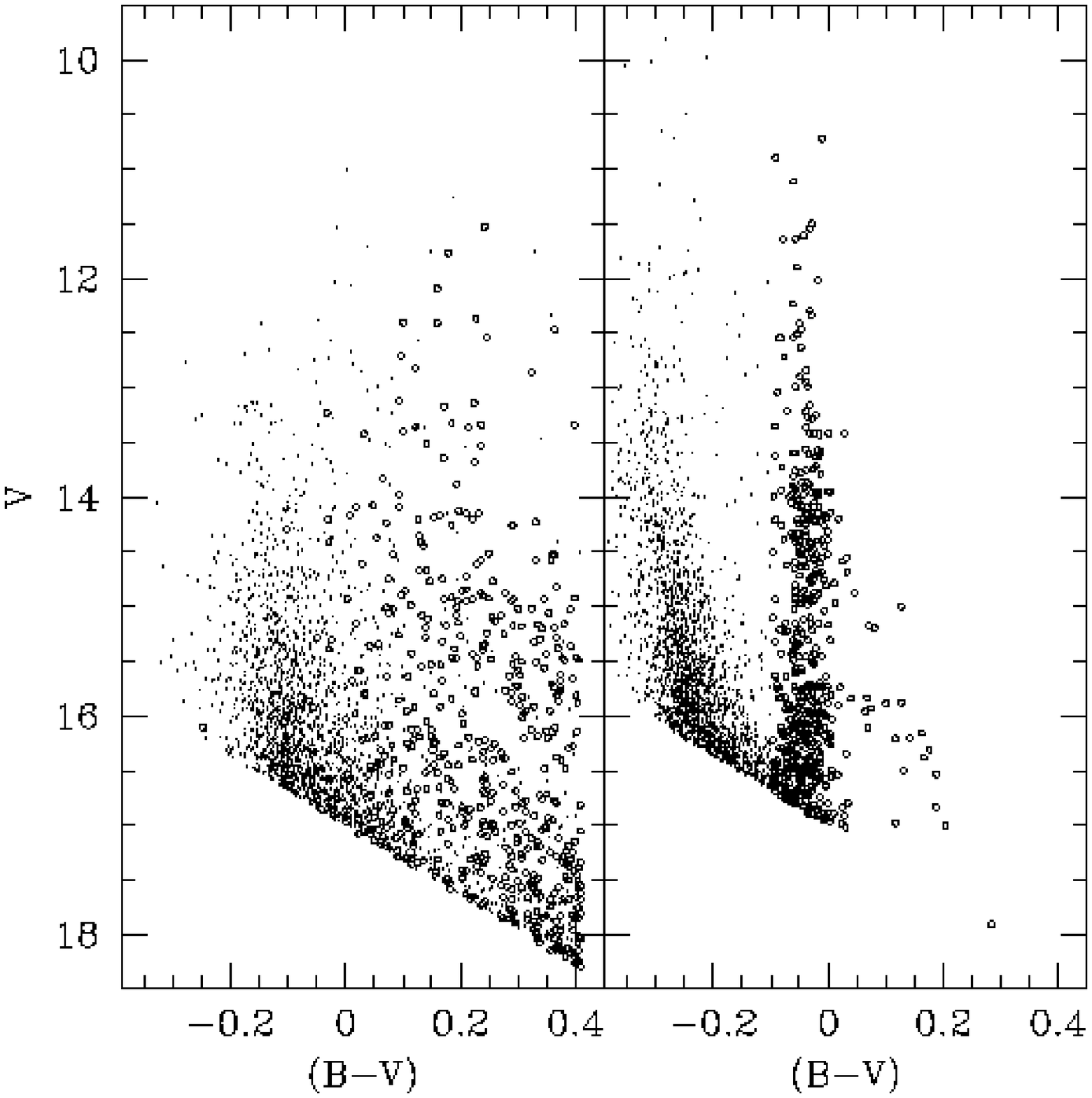}
\end{figure}

\begin{figure}[t]
\plotone{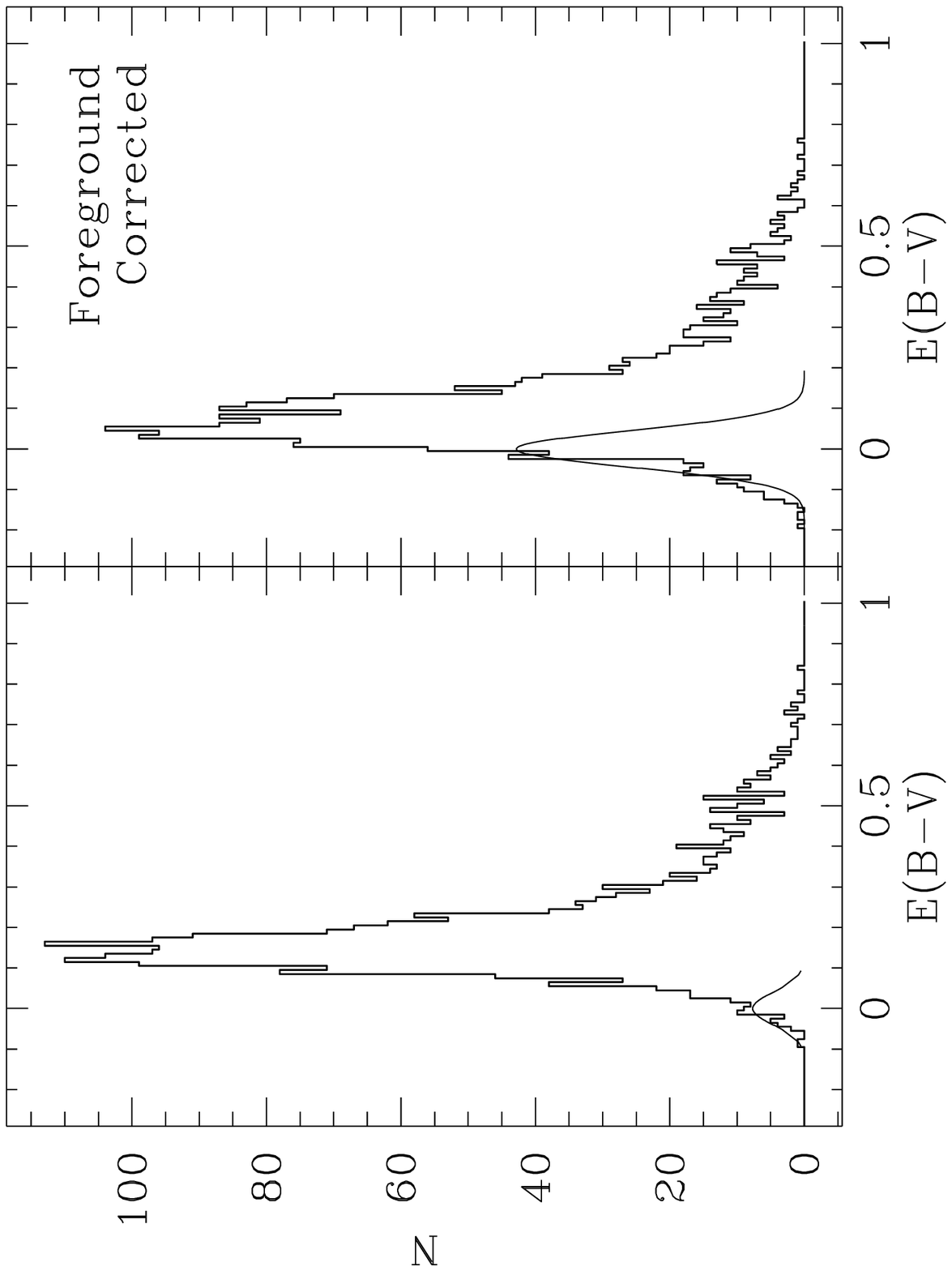}
\end{figure}

\begin{figure}[t]
\plotone{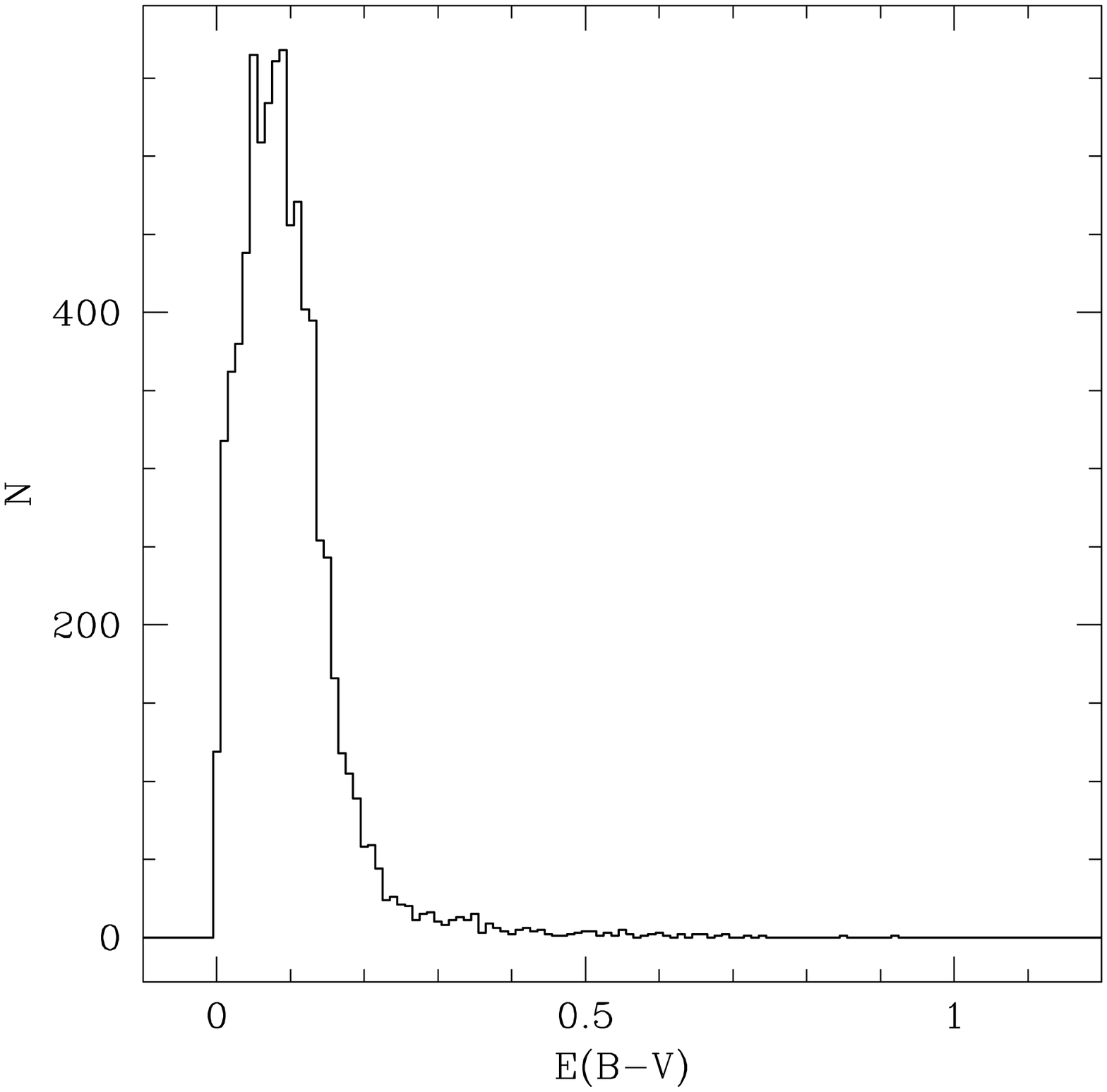}
\end{figure}

\begin{figure}[t]
\plotone{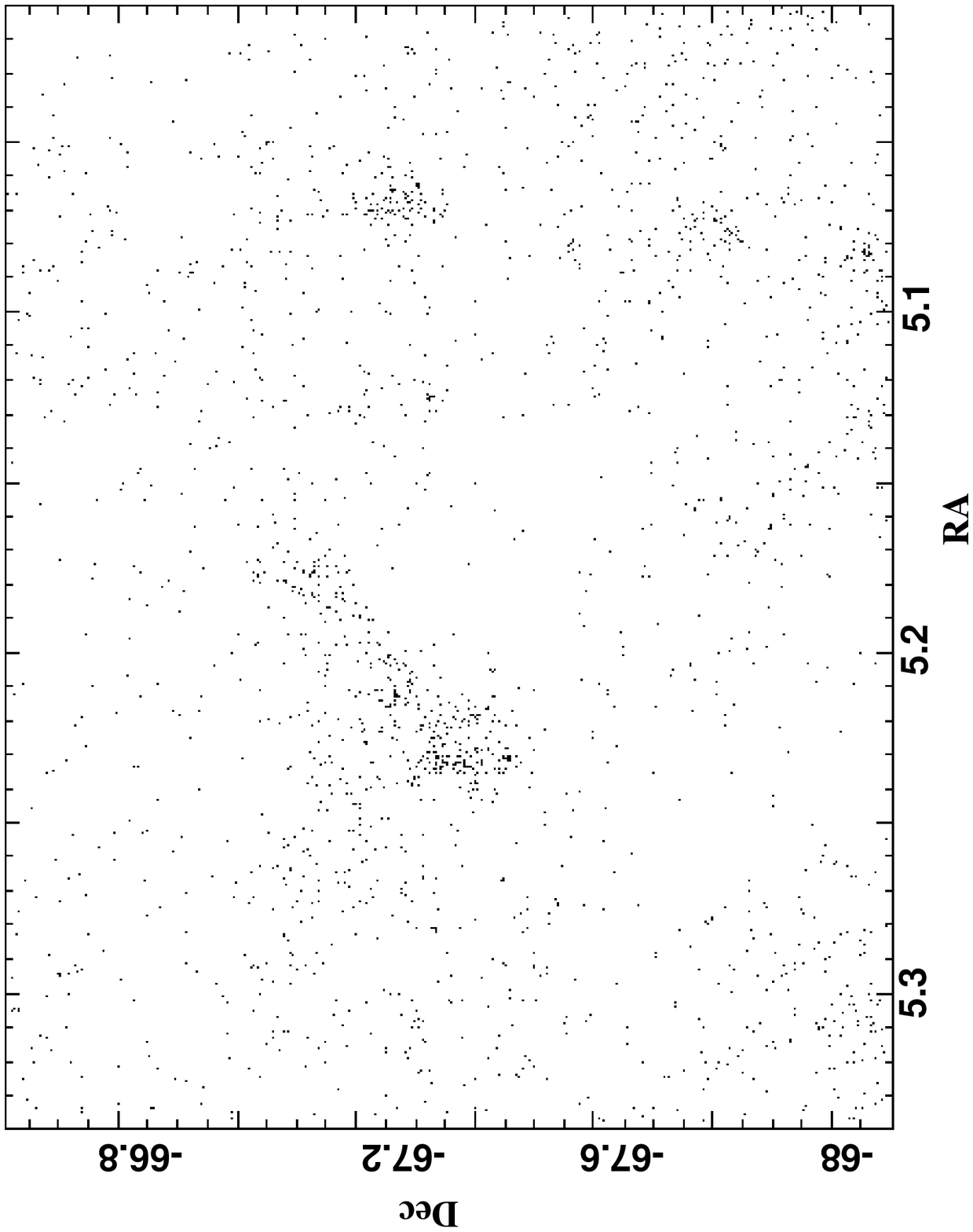}
\end{figure}

\begin{figure}[t]
\plotone{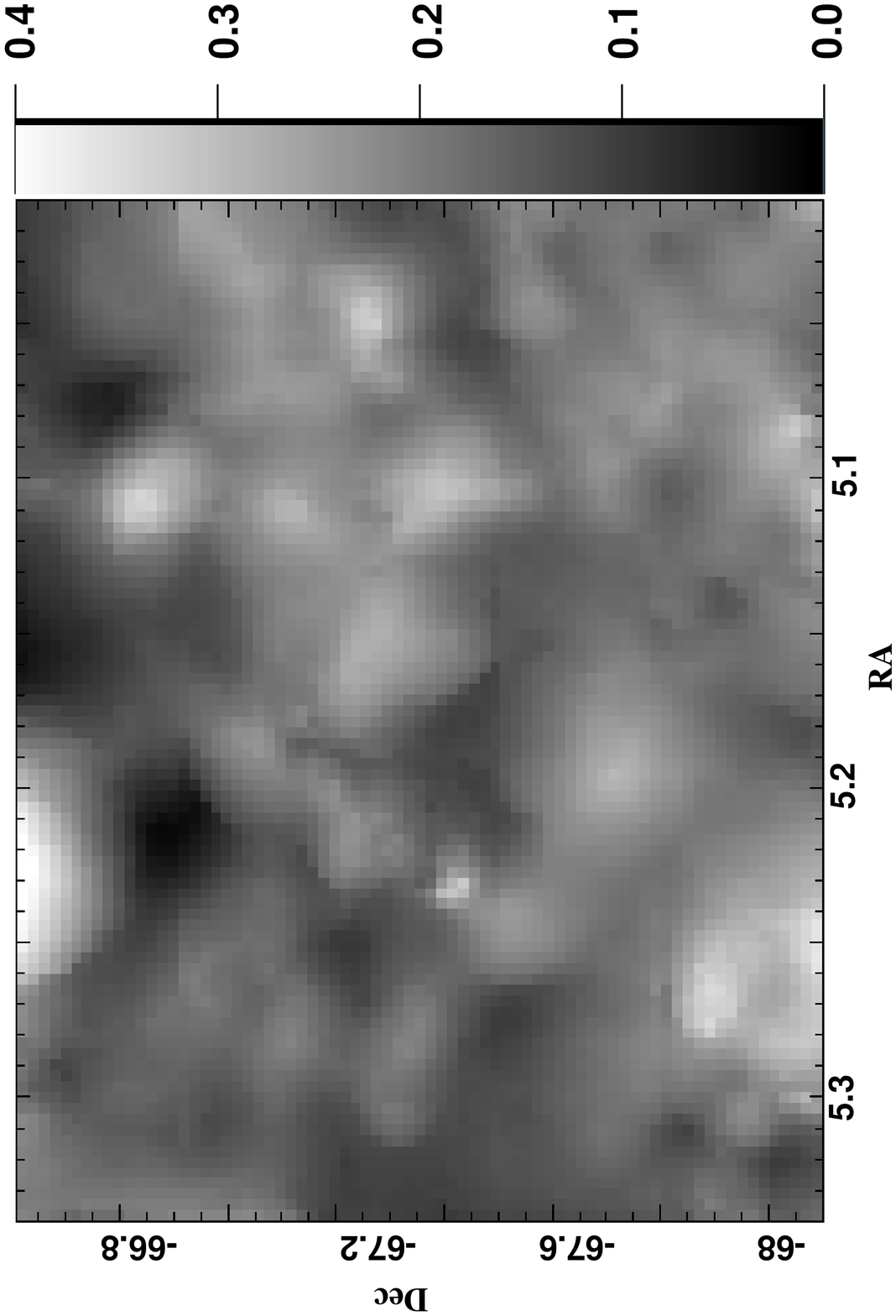}
\end{figure}

\begin{figure}[t]
\plotone{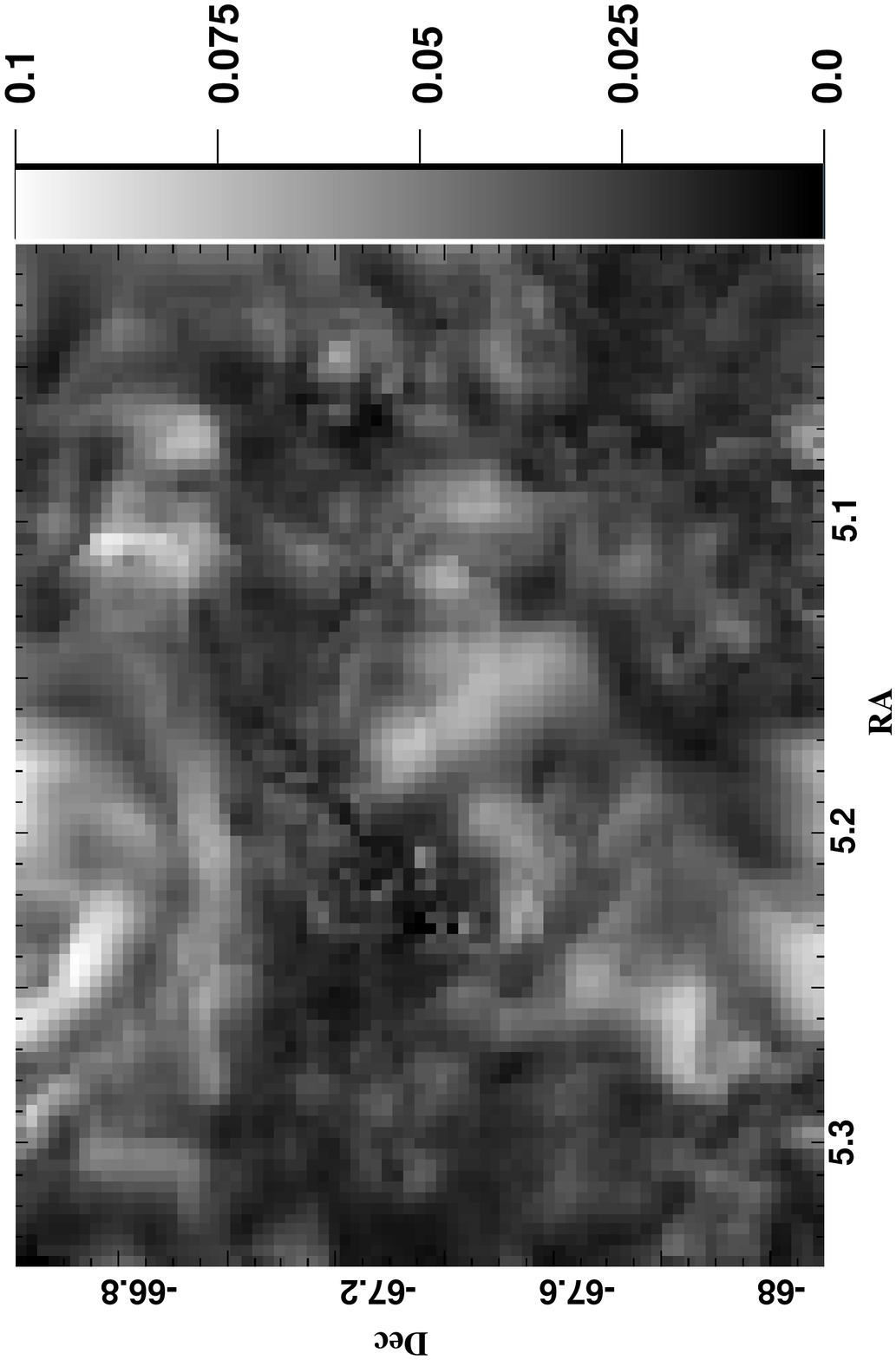}
\end{figure}

\begin{figure}[t]
\plotone{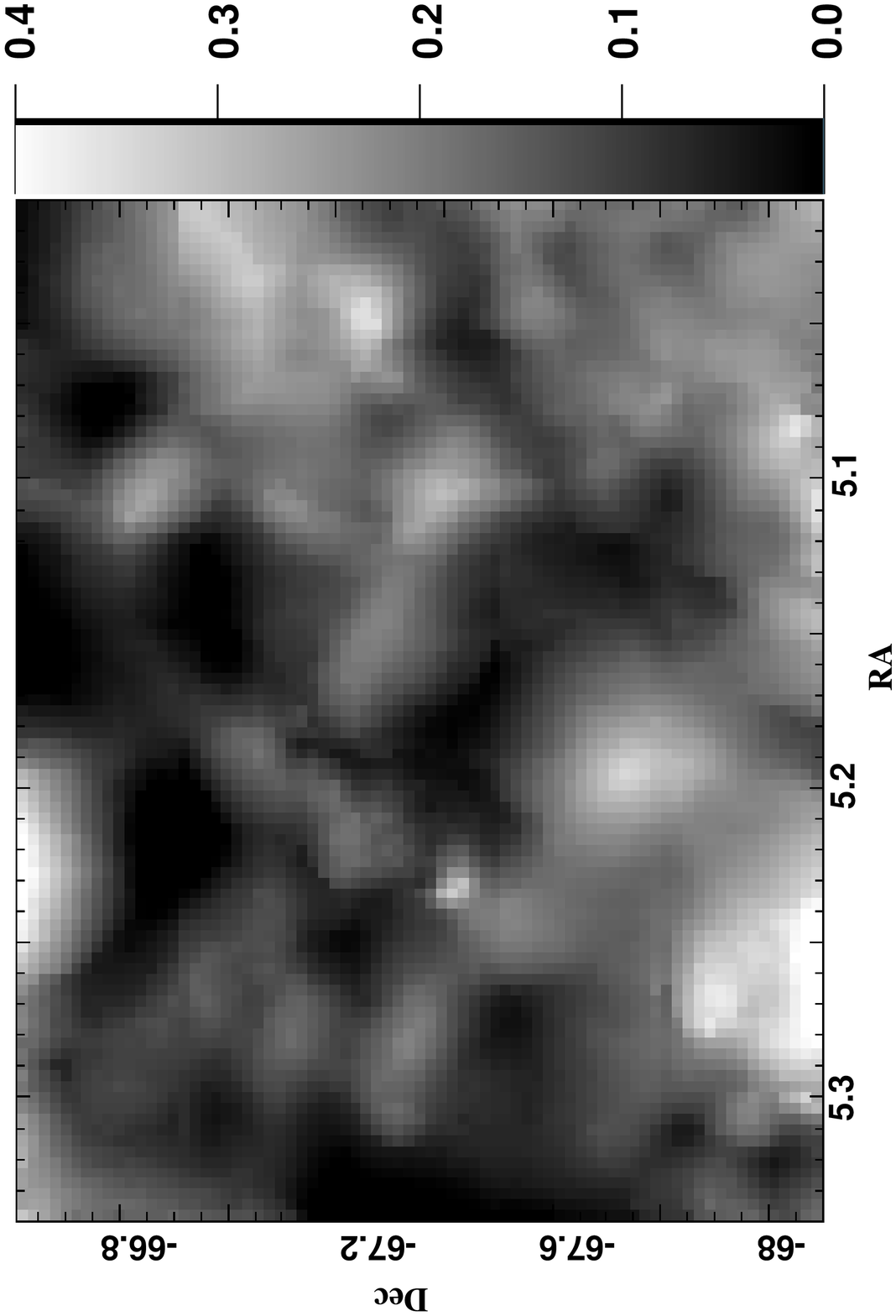}
\end{figure}

\begin{figure}[t]
\plotone{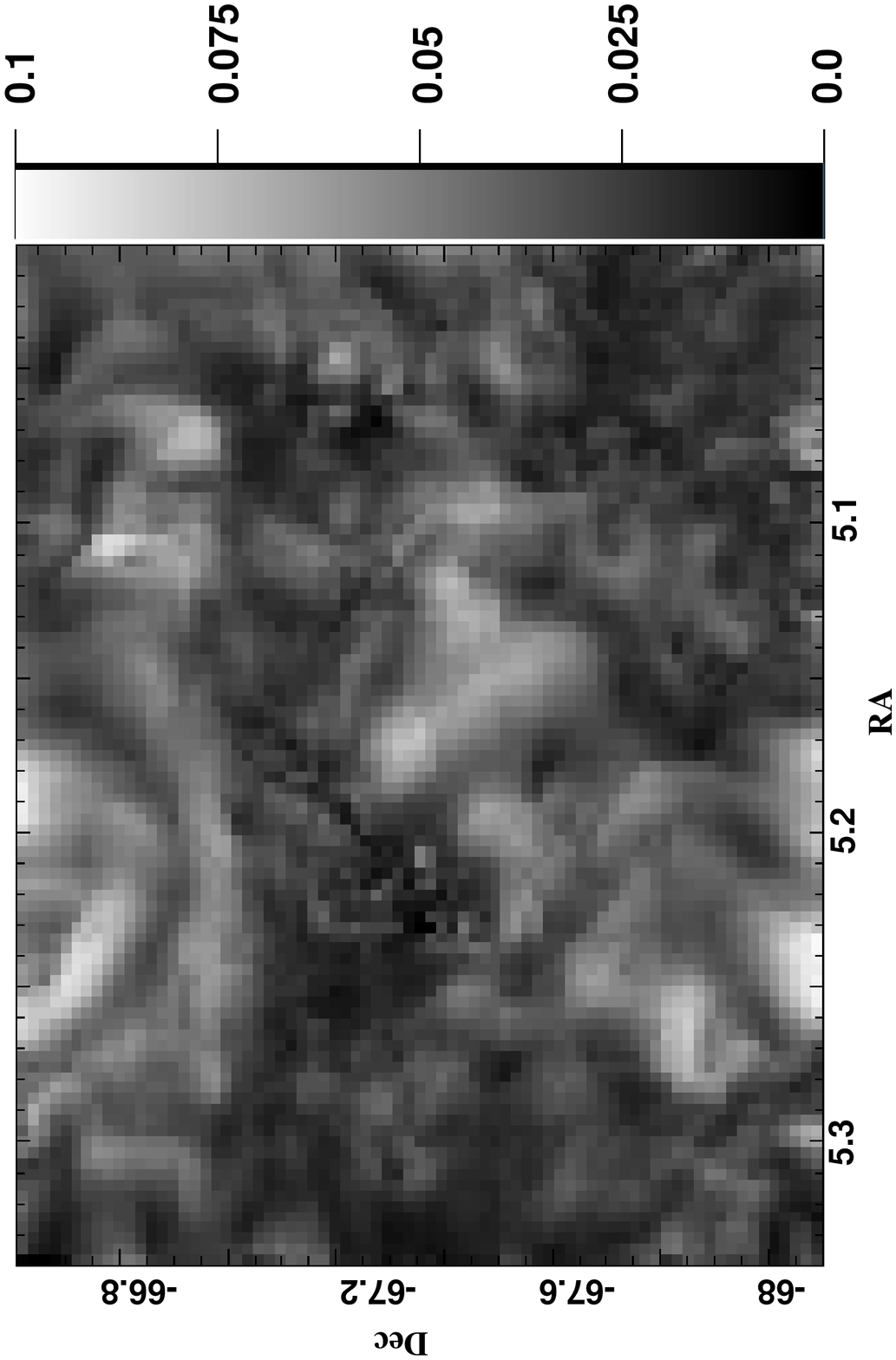}
\end{figure}

\begin{figure}[t]
\plotone{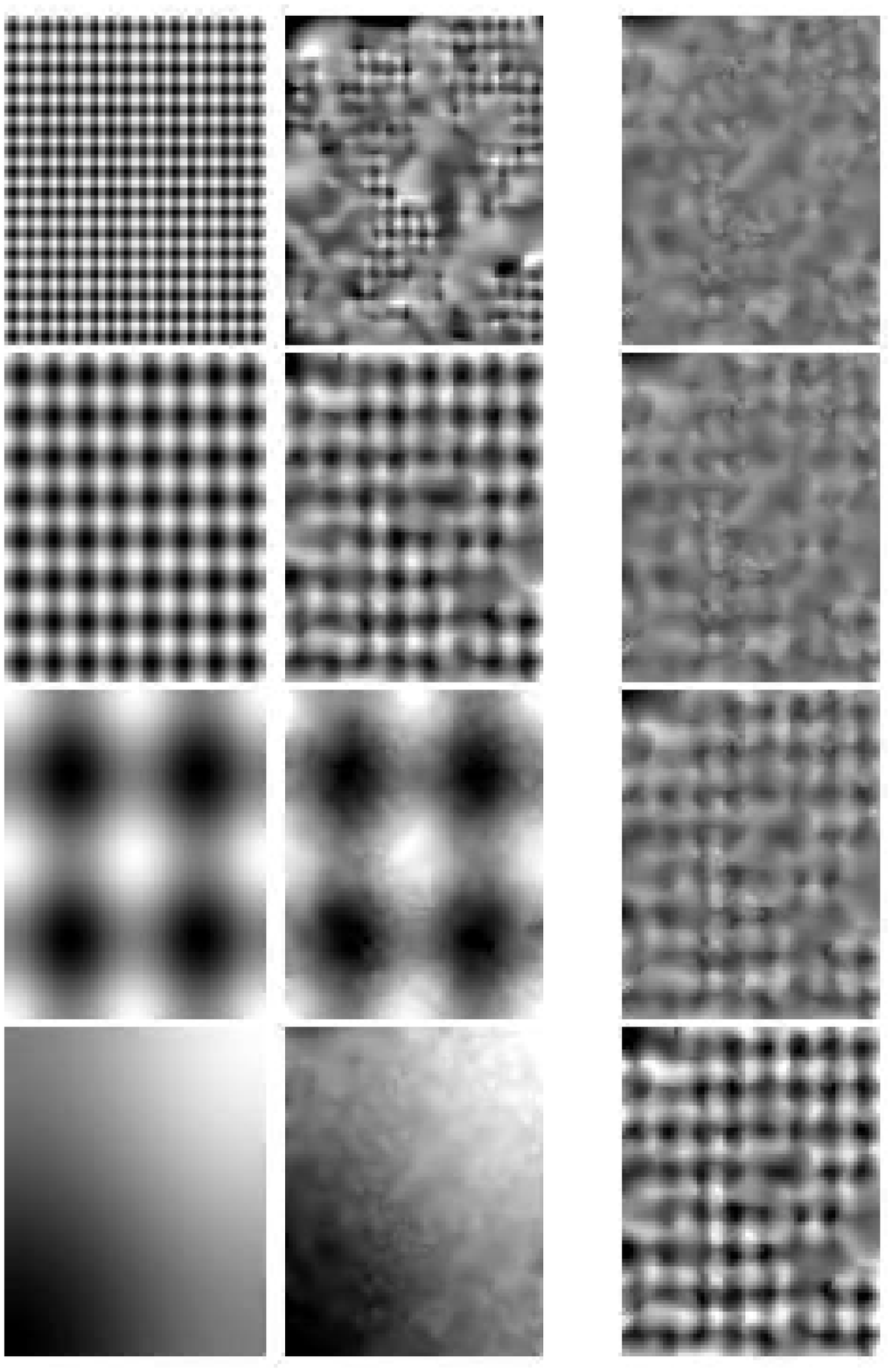}
\end{figure}

\begin{figure}[t]
\plotone{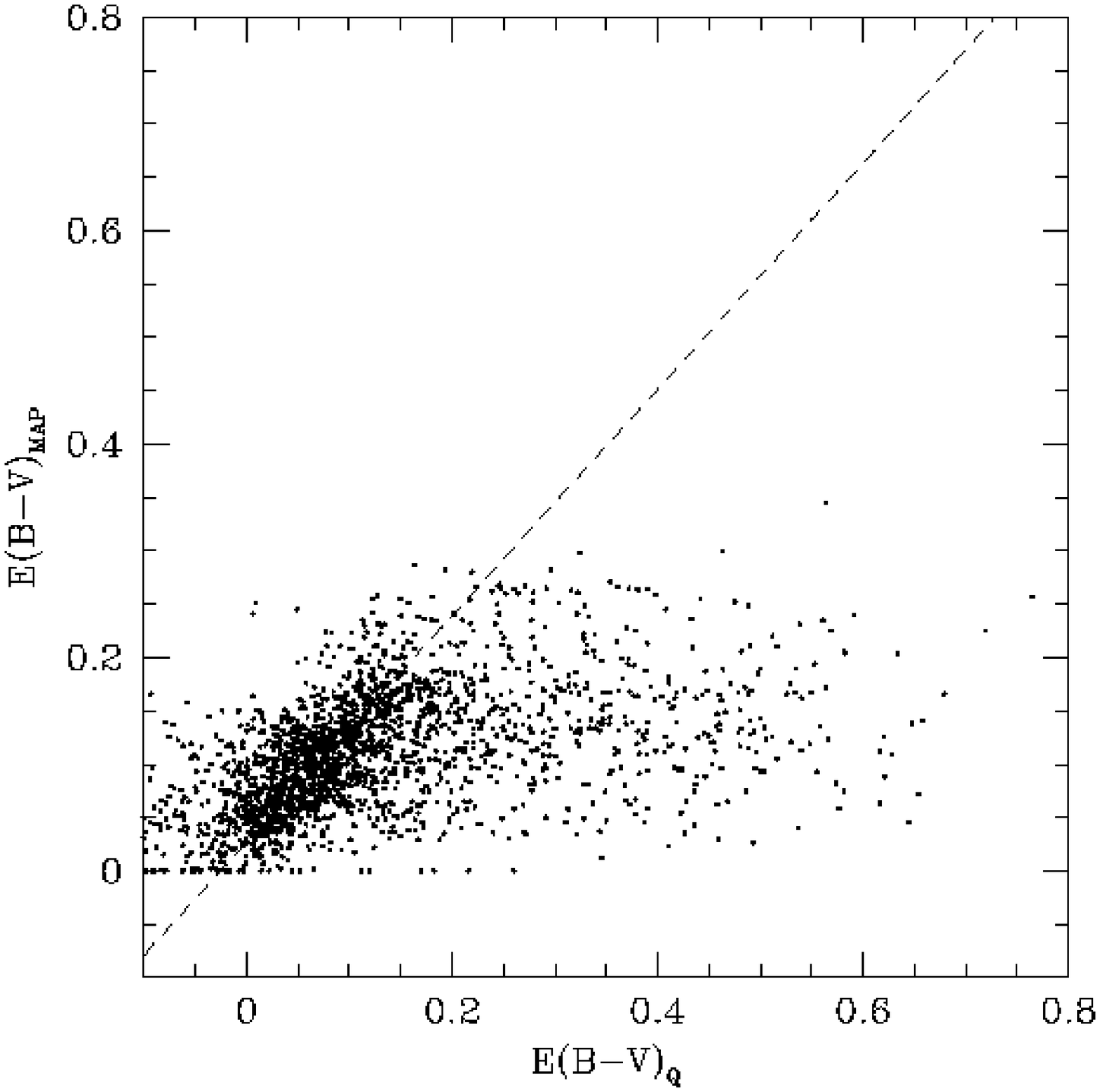}
\end{figure}

\begin{figure}[t]
\plotone{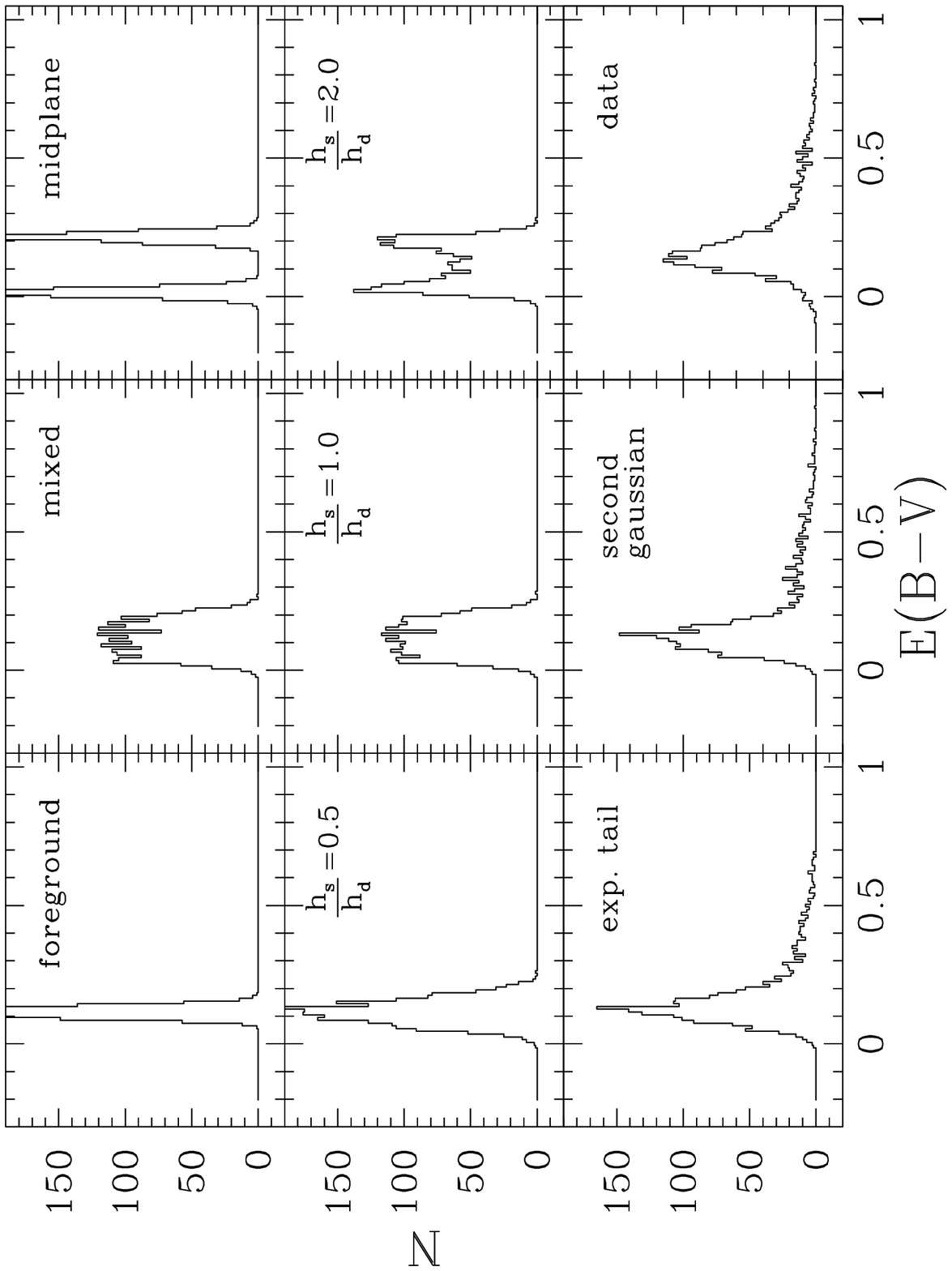}
\end{figure}

\begin{figure}[t]
\plotone{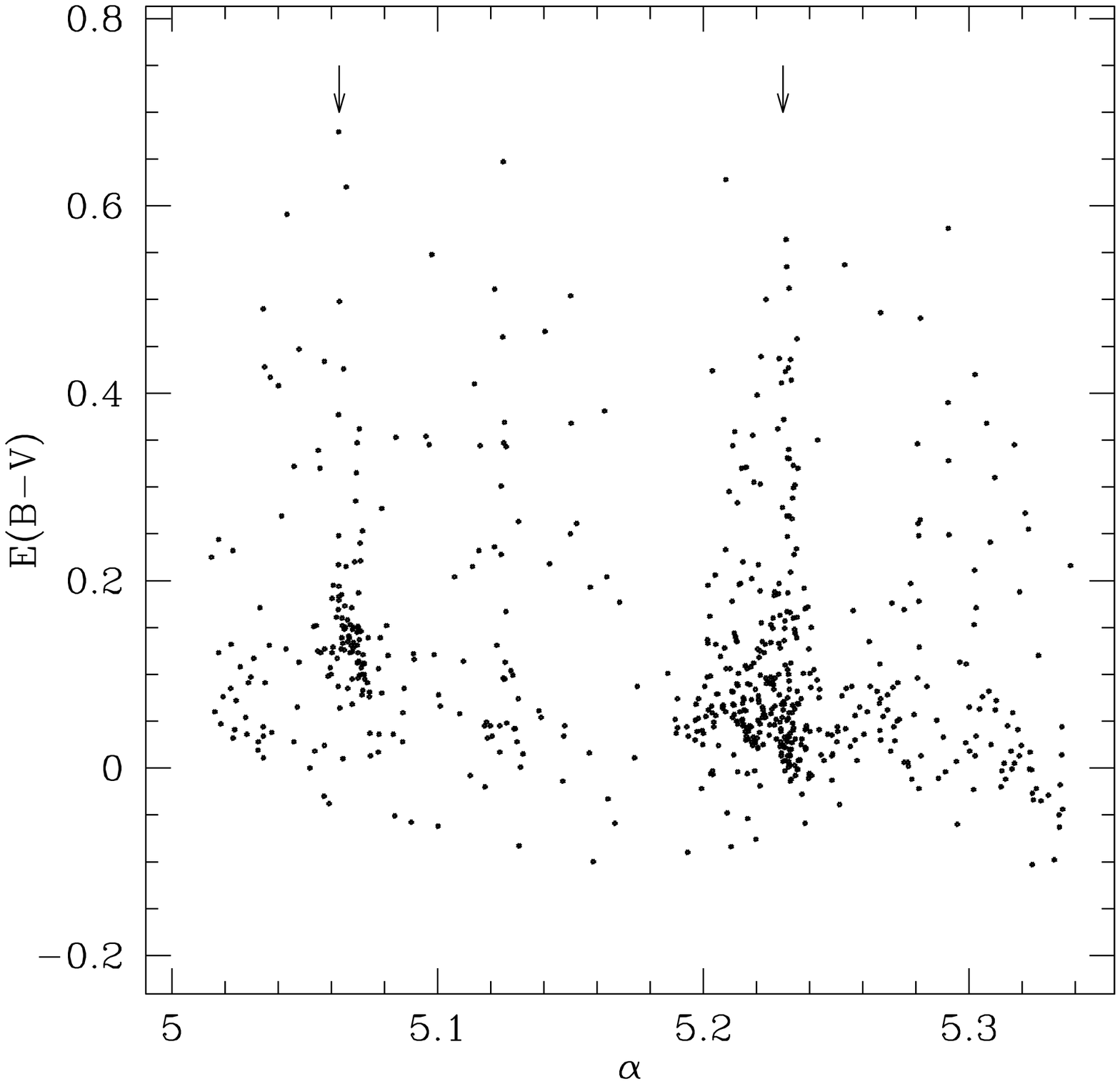}
\end{figure}

\begin{figure}[t]
\plotone{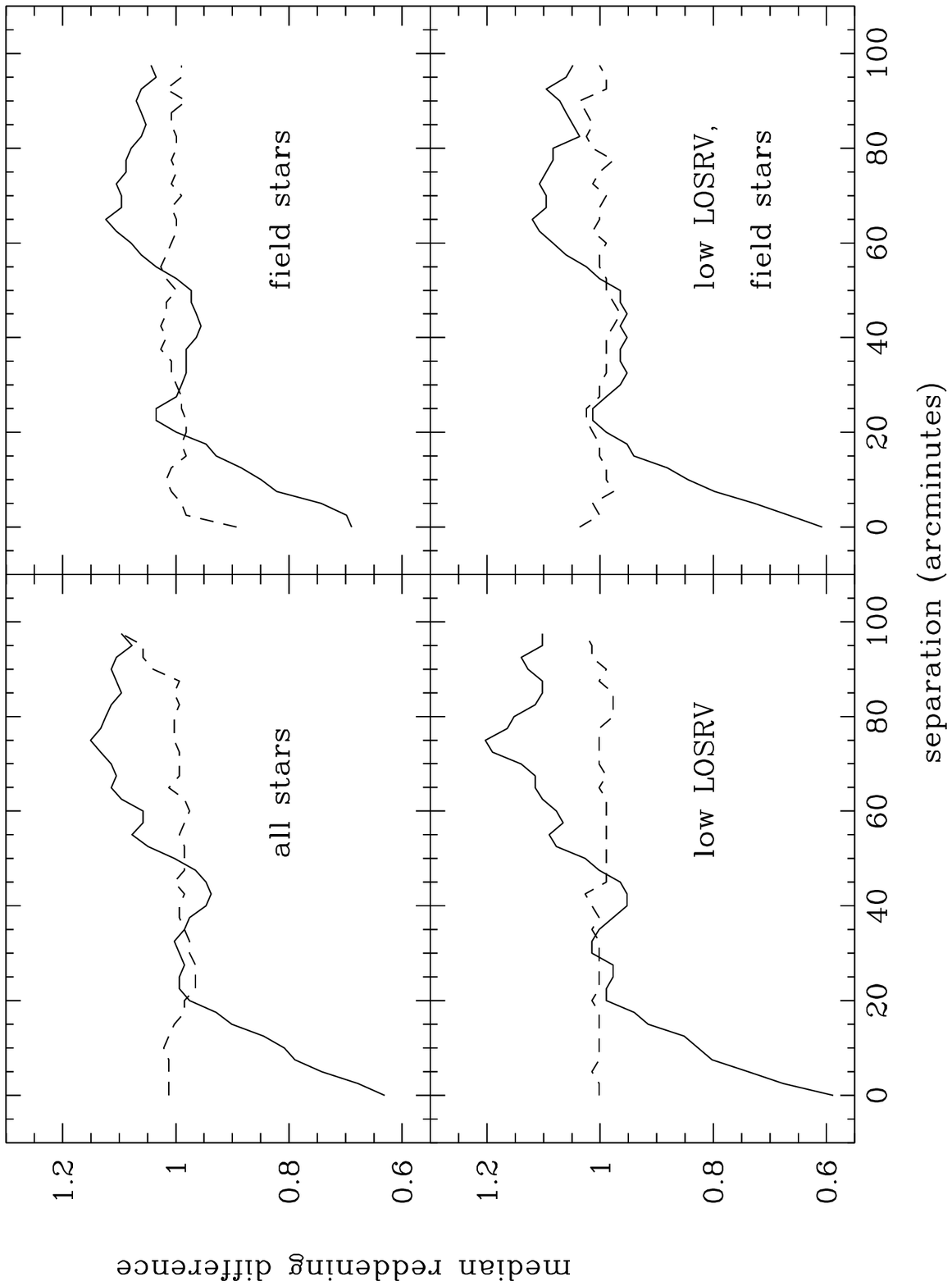}
\end{figure}

\begin{figure}[t]
\plotone{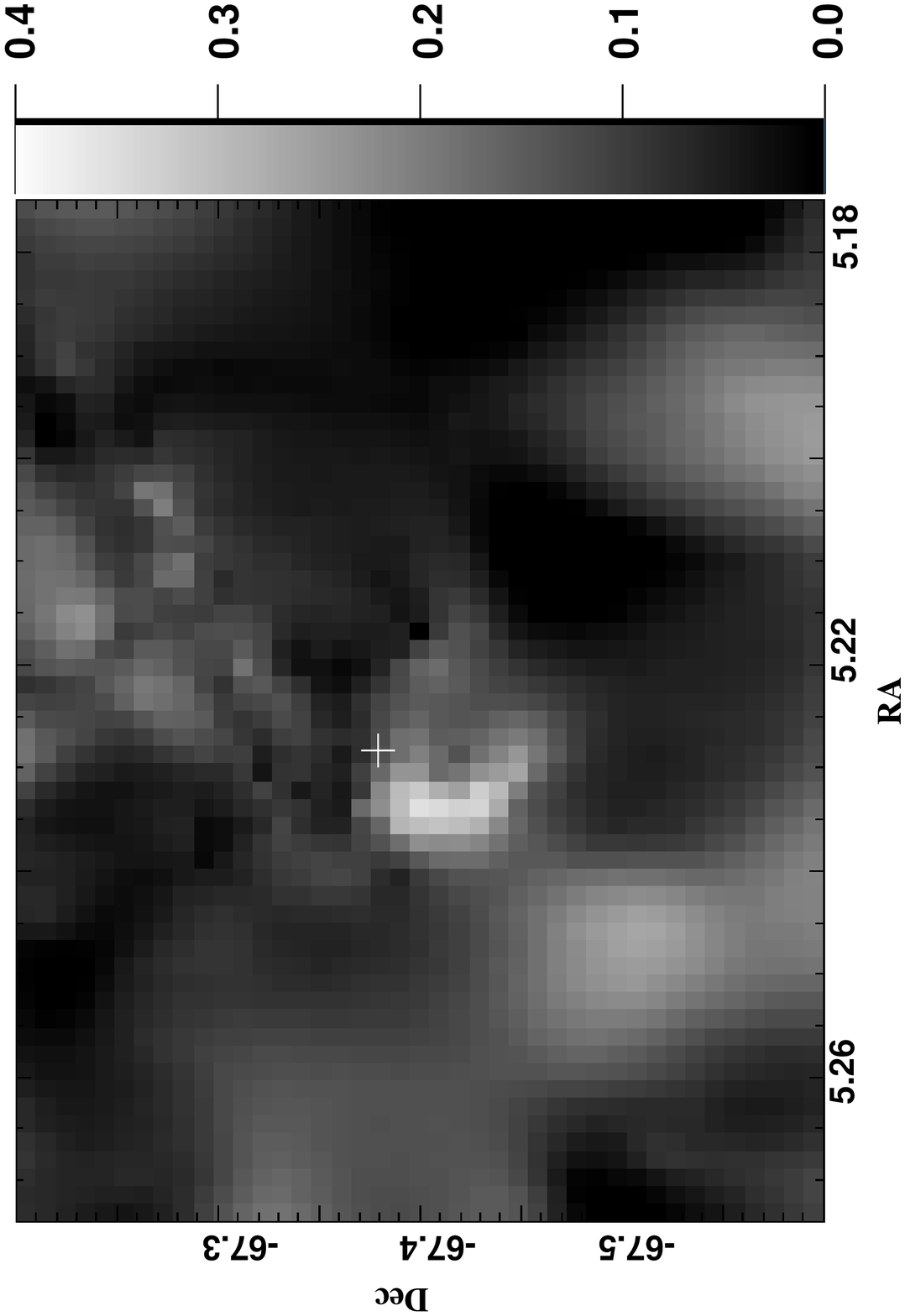}
\end{figure}

\end{document}